\documentclass[%
 reprint,
 amsmath,amssymb,
 aps,
 prl,
]{revtex4-2}

\usepackage{graphicx}
\usepackage{dcolumn}
\usepackage{bm}

\usepackage{textgreek}
\usepackage{color}
\newcommand{\dt}{\mathrm{d}t}
\newcommand{\ind}{\mathfrak{L}}

\begin{document}

\preprint{APS/123-QED}

\title{Dynamical Generation of Epsilon-Near-Zero Behaviour via Tracking and Feedback Control}

\author{Jacob Masur}
 \email{jacobmasur1@gmail.com}
 \author{Denys I. Bondar}
 \email{dbondar@tulane.edu}
\author{Gerard McCaul}
 \email{gmccaul@tulane.edu}
\affiliation{Tulane University Department of Physics and Engineering Physics}

\date{\today}

\begin{abstract}
To date, epsilon near zero (ENZ) responses, characterized by an infinite phase velocity, are primarily achieved by applying a monochromatic light source to a tailored metamaterial. Here, we derive the equations for inducing a dynamically generated broadband ENZ response in a large class of many-body systems via tracking and feedback control. We further find that this response leads to a current-energy relationship identical to that of an ideal inductor. Using a Fermi-Hubbard model, we numerically confirm these results which have the potential to advance optical computation on the nanoscale.
\end{abstract}
             
\maketitle

\paragraph{Introduction.} One of the principal research goals of the twenty first century optics has been the development and application of epsilon-near-zero (ENZ) materials. These form a subclass of near zero index (NZI) materials, whose common feature is the extraordinary optical properties they possess \cite{WuENZPhotonics, ziolkowski2004propagation, ENZPhotonics, nziPhotonics, riseNZItech, nziPhotonicMaterials}. Such materials exhibit a near zero permittivity or permeability at a given frequency, leading to a decoupling of the electric and magnetic fields \cite{ENZPhotonics, nziPhotonics, engheta2006metamaterials, ziolkowski2004propagation, engheta2013pursuing}. Such behaviour has exceptional potential to further our capabilities in the generation and manipulation of nonlinear optical effects at the nanoscale \cite{Vincenti:20}, and represent a vital element in many proposed quantum technologies. For example, ENZ materials can be used to control the phase and direction of emission in antennas \cite{Tong_2020, vlasov2001silicon, PhysRevLett.89.213902}, while in optical circuits, the can behave effectively as components such as switches \cite{https://doi.org/10.1002/adma.201700754, 9144401, bohn2021all, kuttruff2020ultrafast, ZHANG2020106271, Zhang_2019} and modulators \cite{6194256, 6967699, 7386561, Baek:15, lee2014nanoscale, Vasudev:13, liu2018acs}.

To date, experimental realization of these materials is primarily achieved through the construction of artificially constructed metamaterials \cite{doi:10.1063/1.3665414, maasVisibleENZ, zhao2019novel, kelley2019Multiple, Dai:15, CaligiuriPaleiBiffiKrahne, PhysRevB.101.165301, Koivurova_2020, LEE2021412598} and at the plasma frequency in materials with a Drude dispersion relation \cite{anderegg1971optically, spitzer1959infrared, korobkin2006enhanced, caldwell2015low, kim2016role, naik2011oxides, naik2013alternative, kinsey2015epsilon, ou2014ultraviolet}. Though convenient and compatible with complementary metal oxide semiconductor (CMOS) devices \cite{WuENZPhotonics}, there are large intrinsic loses in the metals and semiconductors in which irradiation at the plasma frequency induces an ENZ response \cite{ENZPhotonics}. This is precisely a consequence of the definitional property of an ENZ, namely its near-zero dielectric permittivity \cite{PhysRevLett.117.107404}. Moreover, the bandwidth of current metamaterials is narrow, limiting ENZ metamaterials' usefulness in real-world applications \cite{avignon2017negative, youla1959bounded, tretyakov2001meta, 4231250}. Recent work has shown however that coupling metamaterials to circuits with negative reactance, so-called non-Foster circuits \cite{montgomery1987principles}, can increase their bandwidth and ameliorate this issue \cite{avignon2017negative, hrabarnegativecapacitor, hrabar2010towards}.

Given the current limitations of metamaterial design and fabrication, one might ask if there is an alternative route to achieving the sought-for optical properties exhibited by an ENZ material. Since this behaviour characterises the response of a material to optical driving, an alternative to engineering the equilibrium properties of materials might be to instead consider how the relationship between driving and response can characterised and manipulated. Specifically, we ask whether an ENZ-like response be generated \textit{dynamically} by designing a driving field whose response functionally emulates that expected from an ENZ, namely an infinite phase velocity. 

To achieve this objective, it is necessary to import techniques from the field of quantum control \cite{d2021introduction, dong2010quantum}. In particular, \textit{tracking control} \cite{PhysRevA.72.023416, PhysRevA.98.043429, PhysRevA.84.022326, PhysRevLett.118.083201, doi:10.1063/1.1582847, doi:10.1063/1.477857, Gerard1, Gerard2} provides a method for calculating driving fields such that the evolution of a given observable tracks a specified trajectory. These methods have been deployed profitably in the context of atomic \cite{PhysRevLett.118.083201}, molecular \cite{PhysRevA.98.043429,Magann2022,Magann2023} and solid-state systems \cite{Gerard1,Gerard2,twinning_fields,nonuniqueness}. Complementary to this method is \textit{feedback control}  \cite{aastrom2021feedback}. This has been implemented in a wide variety of systems for applications ranging from cruise control \cite{powers2000automotive, kiencke2000automotive, barron1996role} to computer network protocols \cite{low2002internet, tanenbaum1981network, jacobson1988congestion, hellerstein2004feedback} and atomic force microscopy \cite{sarid1991atomic, schitter2001high}. Recently the concept of feedback control has been further extended into the realm of quantum algorithm optimisation \cite{PhysRevA.106.062414,PhysRevLett.129.250502}, but to date has not been applied to the manipulation of quantum system's nonlinear optical properties.  

In this letter, we demonstrate that using either tracking or feedback control allows for the dynamical generation of an ENZ response in a many-body system. Furthermore, we show that this behaviour mimics that of an ideal inductor, raising the possibility that feedback control mechanisms can be used to create optical analogues to the fundamental components of electrical circuitry. Significantly, the proposed method admits the possibility of negative inductances, furnishing precisely the type of non-Foster behaviour necessary for increasing the bandwidth of metamaterials. 

\paragraph{Tracking Control.} We begin by outlining the procedure for tracking control, and detail how this is applied to generate an ENZ response in a many-body system. We consider a one-dimensional model throughout this letter for the sake of simplicity, but the following analysis can be generalized to higher dimensions. We take as a model a general many-body fermionic system in which $N_e$ electrons are localized to lattice sites $j = 1, ..., N_s$ with spin $\sigma = \uparrow, \downarrow$. The kinetic energy is described by the hopping parameter $t_0$, acting as an effective single band system coupled to a driving laser via the dipole approximation. The effect of this is to introduce a phase $\exp[\pm i\Phi(t)]$ onto the hopping parameter, where $\Phi(t)$ is the Peierls phase \cite{Peierls1933,PhysRevA.95.023601}. The Hamiltonian is then described by (in atomic units):
\begin{equation} \label{general_ham}
	\hat{H}(t) = -t_0 \sum_{j,\sigma} \left(e^{-i\Phi(t)} \hat{c}^\dag_{j, \sigma} \hat{c}_{j+1, \sigma} + \mathrm{h.c.}\right) + \hat{U}
\end{equation}
where  $\hat{c}^{\dag}_{j,\sigma}$ and $\hat{c}_{j,\sigma}$ are, respectively, the canonical fermionic creation and annihilation operators, and $\hat{U}$ describes the potential resulting from electron-electron repulsion.

Optical driving induces a current in the system which is calculated by the expectation of the current operator. The current operator is derived by a continuity equation \cite{Gerard1,Gerard2,twinning_fields}, giving:
\begin{equation} \label{currentop}
	\hat{J}(t) = -iat_0 \sum_{j, \sigma} \left(e^{-i\Phi(t)} \hat{c}^\dag_{j, \sigma} \hat{c}_{j+1, \sigma} - \mathrm{h.c.}\right)
\end{equation}
where $a$ is the lattice constant. The optical emission of a driven material is determined exclusively by this current, and forms the basis for the study of optical response in the solid state \cite{HHGreview,Ghimire2018,McDonald_2015,Silva2018}. This is due to the fact that under the dipole approximation, the emitted field is proportional to the dipole acceleration $\frac{d J(t)}{dt}$, where $J(t)\equiv \langle \psi(t) | \hat{J}(t) | \psi(t) \rangle$. For this reason, it is via the control of this observable that an ENZ response can be generated. 

The condition for an ENZ response can be formulated in terms of the relationship between the driving field and response. The infinite phase velocity implied by a near-zero dielectric permittivity implies a zero lag between the applied and emitted electrical fields, i.e. $E_{\mathrm{out}}(t) \propto E_{\mathrm{in}} (t)$. This relationship can be expressed in terms of $\Phi(t)$ and $J$ by recalling that $E_{\mathrm{in}}(t) = -\frac{1}{a} \frac{d \Phi(t)}{dt}$ \cite{Gerard1}. Thus, the ENZ criteria can be succinctly represented by
\begin{equation} \label{originalcriteria}
    \frac{\mathrm{d}J(t)}{\dt} = -\frac{1}{a\ind} \frac{\mathrm{d} \Phi(t)}{\mathrm{d} t},
\end{equation}
where $\ind$ is a non-zero real constant. Integrating both sides with respect to time, we obtain a condition for an ENZ response $J_{\rm ENZ}$,
\begin{equation} \label{enzcriteria}
    J_{\rm ENZ}(t) = -\frac{1}{a\ind} \Phi_{\rm ENZ}(t) + C
\end{equation}
which can be implemented directly in a tracking control framework.

The laser field $-\frac{1}{a} d\Phi_{\rm ENZ}(t)/dt$ that induces an ENZ response is given by the solution to the implicit equation
\begin{equation} \label{enzphi}
    \begin{split}
        \frac{1}{a\ind} \Phi_{\rm ENZ}(\psi) = 2at_0 R(\psi) \sin\left[ \Phi_{\rm ENZ}(\psi) - \theta(\psi) \right]+ J(0).
    \end{split}
\end{equation}
Here, $J(0)$ is the current expectation of the initial state, and $R(\psi)$ and $\theta(\psi)$ are, respectively, the real and imaginary components of the nearest neighbor expectation $\langle \sum_{j, \sigma} \hat{c}^\dag_{j, \sigma} \hat{c}_{j+1, \sigma} \rangle$. The derivation of Eq.~\eqref{enzphi} is provided in Sec. I of the supplemental materials, and the existence of such a field and the conditions under which it is unique are demonstrated in Sec. II \cite{supplement}. Finally, it is worth noting that the physical validity of this tracking equation in the $\dt\rightarrow 0$ limit can be checked by comparison to the Ehrenfest theorem for $J(t)$ (see e.g. \cite{Gerard2}).

\begin{figure*}
  \includegraphics[width=\textwidth]{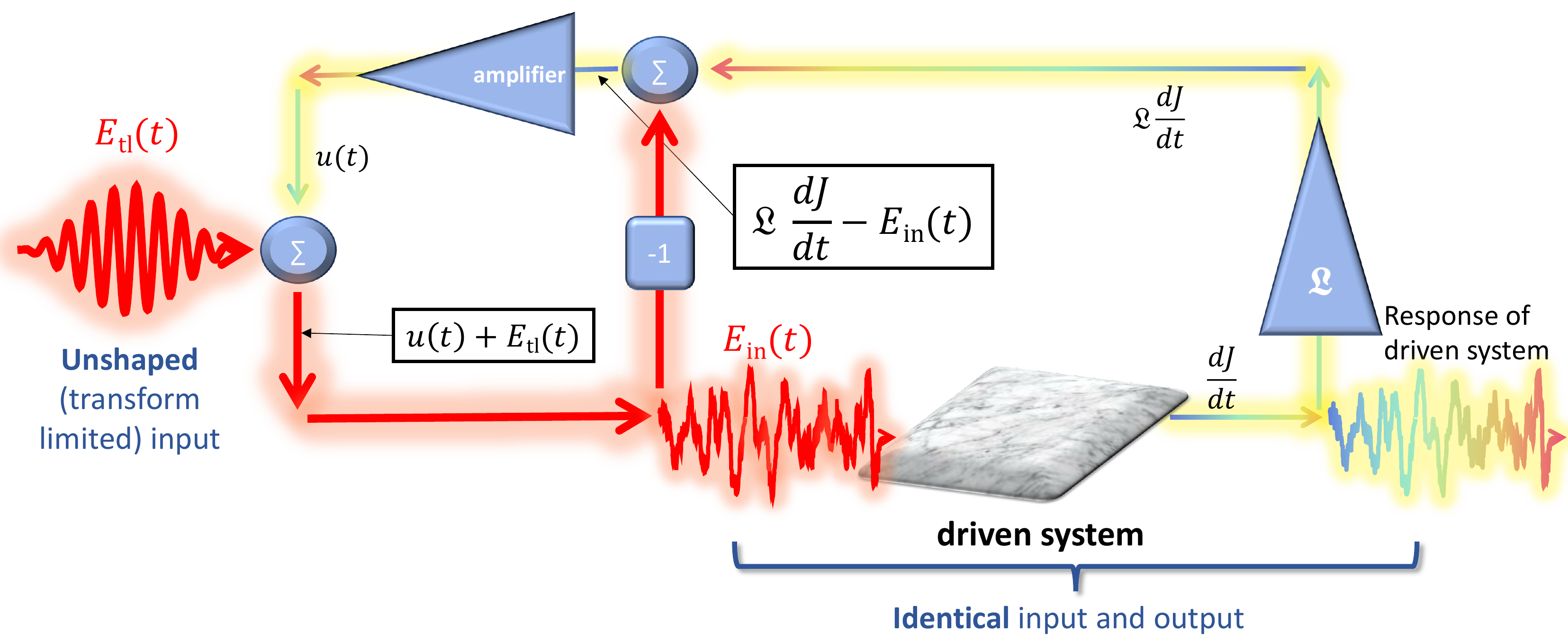}
  \caption{The schematic depicting the setup of ENZ feedback control. The response of the system is amplified or dampened by some value proportional to $\ind$, from this the input field is subtracted, and the resulting field is amplified by $k_p$ before being combined with the transform limited pulse.}
  \label{fig:feedback_control}
\end{figure*}

\paragraph{Feedback Control.} An alternative route to obtaining an ENZ response can be obtained by using the material's own response to control the input field. Here, we implement a simple form of feedback control known as proportional feedback control, a schematic for which is given in Fig.~\ref{fig:feedback_control}. In this scenario,  we begin by applying a transform limited pulse with electric field $E_{\rm tl} (t)$ which is subsequently modified by the introduction of the control field
\begin{equation} \label{control_field}
    u(t) = k_p\left( \ind \frac{\mathrm{d} J(t)}{\dt} - E_{\rm in}(t) \right)
\end{equation}
where $k_p$ is a positive constant representing the amplification gain. The control field enforces the ENZ condition \eqref{originalcriteria} by correcting the input field when the error $e = \ind \frac{\mathrm{d} J}{dt} - E_{\rm in}$ is nonzero.

The combined input to the system when we correct the transform limited field with the control field \eqref{control_field} is then
\begin{equation} \label{input_field}
    E_{\rm in}(t) = E_{\rm tl}(t) + u(t).
\end{equation}
Substituting $E_{\rm in}(t)$ as given above into Eq.~\eqref{control_field} together with some algebraic rearrangement allows us to express it as
\begin{equation}
    u(t) = \frac{k_p}{1 + k_p} \left( \ind \frac{\mathrm{d}J(t)}{\dt} - E_{\rm tl}(t) \right),
\end{equation}
which in the limit $k_p \rightarrow \infty$, gives $u(t) = \ind \frac{\mathrm{d}J(t)}{\dt} - E_{\rm tl}(t)$. Then, by Eq. \eqref{input_field}, the input field approaches
\begin{equation}
    E_{\rm in}(t)\to E_{\rm ENZ}(t)=\ind \frac{\mathrm{d}J(t)}{\dt},
\end{equation}
the ENZ condition. Hence, \emph{given sufficiently strong amplification $k_p$, the feedback scheme Fig.~\ref{fig:feedback_control} will drive any system to produce ENZ response. }

To computationally illustrate this, we calculate the Peierls phase by multiplying Eq. \eqref{input_field} by $-a$ and integrating over time:
\begin{equation} \label{feedpack_phi}
  \Phi_{\rm in} (t) = \Phi_{\rm tl}(t) + \frac{k_p}{1 + k_p} \left[ -a\ind \left(J(t) - J(0)\right) - \Phi_{\rm tl}(t) \right].
\end{equation}

\paragraph{Physical Interpretation of $\ind$.} A natural question is how the free parameter $\ind$ should be interpreted. When considering Eq.~\eqref{originalcriteria}, it bears a strong resemblance to the dynamics of an inductor  \cite{pollack2002electromagnetism}. For an inductor with intrinsic inductance $L$ and length $d$ subject to a homogeneous time dependent electric field $E(t)$, we have 
\begin{equation}
    \frac{\mathrm{d} I(t)}{\dt} = \frac{d}{L} E(t)
\end{equation}
Here, $I(t)$ is the macroscopic current which varies only in time. The total energy $\mathcal{E}$ stored by an inductor at time $t$ is
\begin{equation}
\label{inductorenergy}
    \mathcal{E}(t) - \mathcal{E}(0) = \frac{L}{2} \left[ I^2(t) - I^2(0) \right].
\end{equation}

In the scenario we consider, the dynamics take place on the nanoscale, so we substitute the macroscopic current for the current expectation $J(x,t) = \langle \hat{J}(x, t) \rangle$. Then, we discretise with $J(x, t) \rightarrow \frac{1}{a} J_j(t)$ and $\int \mathrm{d} x \rightarrow a \sum_{j}$ to obtain the relationship between energy and current in a system defined by Eq.~\eqref{general_ham}. The power stored by the solid-state system will be
\begin{equation}
    P(t) = E(t) \sum_j J_j(t) = E(t) J(t),
\end{equation}
which after time integration yields 
\begin{equation}
    \mathcal{E}(t) - \mathcal{E}(0) = -\frac{1}{a} \int_0^t \frac{{\rm d} \Phi(t')}{{\rm d}t'} J(t') \dt'.
\end{equation}
Upon inserting the specific form of $\Phi(t)$ imposed by the ENZ condition given in Eq.~\eqref{enzcriteria}, we obtain an identical energy-current relationship to that of the inductor shown in Eq.~\eqref{inductorenergy}:
\begin{equation} \label{energycurrentrelationship}
    \mathcal{E}(t) - \mathcal{E}(0) = \frac{\ind}{2} \left[ J^2(t) - J^2(0) \right],
\end{equation}
and hence we are able to identify the value of $\ind$ chosen as an effective inductance.

\paragraph{Simulation results.}
We now demonstrate that both tracking and feedback control induce the desired ENZ response via numerical calculations. Though the Hamiltonian in Eq.~\eqref{general_ham} and subsequent analysis of the field required to induce an ENZ response are valid for any $\hat{U}$ that will commute with electron number operators \cite{twinning_fields,nonuniqueness}, for simulations we take our system to be a half-filled Fermi-Hubbard model ($N_s = N_e$) \cite{PhysRevA.106.013110}, where $N_\uparrow = N_\downarrow$ and onsite interaction of the form
\begin{equation}
    \hat{U} = U \sum_{j=1}^{N} \hat{n}_{j, \uparrow} \hat{n}_{j, \downarrow}
\end{equation}
where $U$ parametrises interaction energy and $\hat{n}_{j, \sigma} = \hat{c}^\dag_{j,\sigma} \hat{c}_{j,\sigma}$ is the number operator for site $j$ and spin $\sigma$. 

Numerical simulations of both types of control are performed using the QuSpin package in Python \cite{quspin}. To avoid trivial solutions of zero field and current, after initialising the simulation in the ground state of the system, we then excite it with 10 cycles of a transform limited pump pulse of strength $F_0 = 10 \frac{\mathrm{MV}}{\mathrm{cm}}$ and frequency $\omega_0 = 32.9$ THz. After this pre-pump, we implement the chosen control procedure. The system is evolved using the DOP853 algorithm \cite{dop853} implemented in SciPy \cite{2020SciPy-NMeth} for a total time equal to the duration of the pump pulse $T = \frac{2\pi M}{\omega_0}$ where $M$ is the number of cycles. At each time step, we solve Eq.~\eqref{enzphi} for tracking control and Eq.~\eqref{feedpack_phi} for feedback control.

\begin{figure}[ht]
\begin{center}
\includegraphics[width=1\columnwidth]{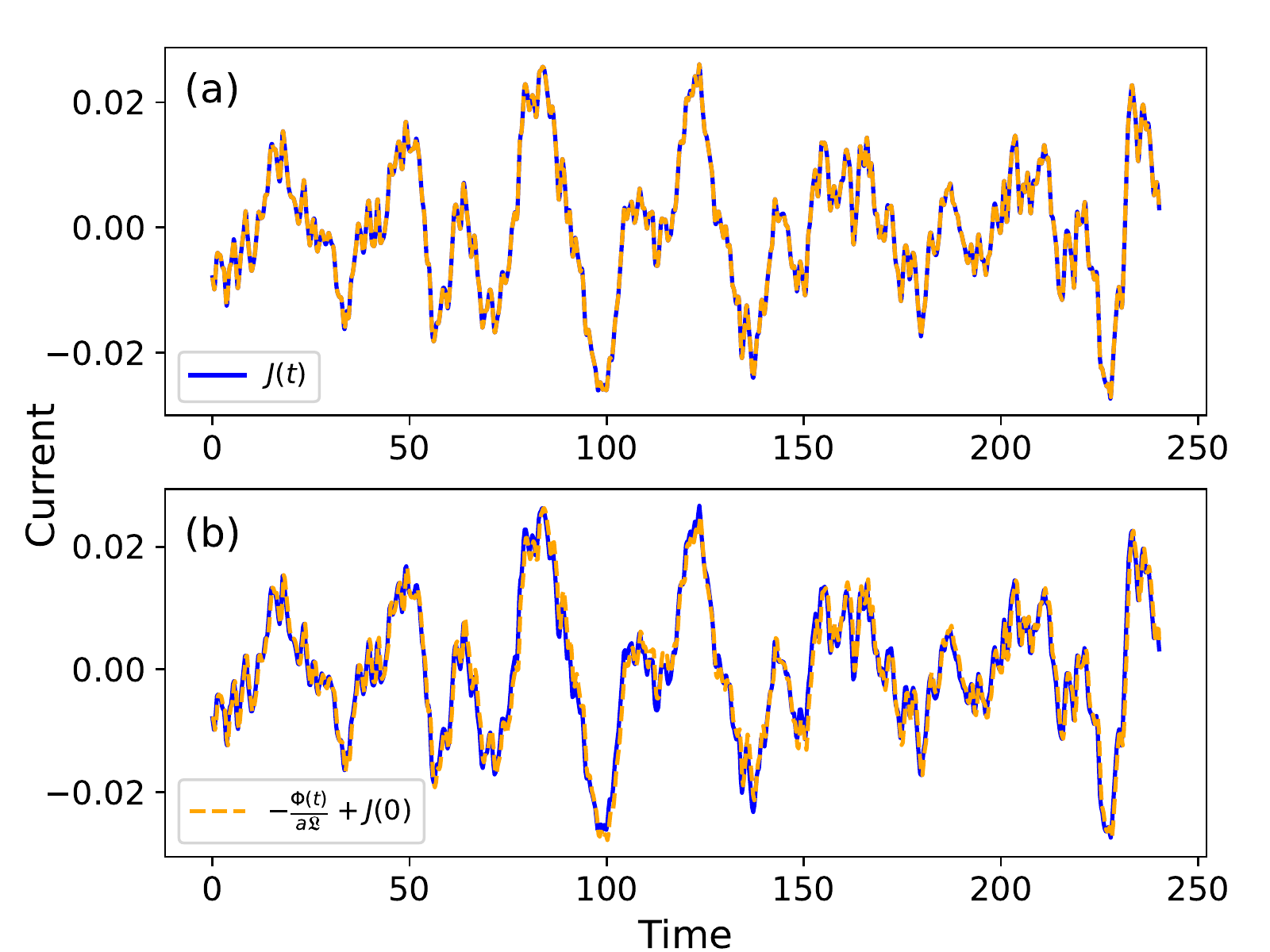}\end{center}
\caption{Plots of the current induced (solid blue line) and $-\frac{\Phi(t)}{a\ind} + J(0)$ (dashed orange line) for Fermi-Hubbard material simulated exactly with 10 sites, $\frac{U}{t_0} = 1$, $a = 4$ \r{A}, and $\ind = 1$. Plot (a) shows the response induced by tracking control and plot (b) shows the response induced by feedback control with $k_p = 10000$.}
\label{fig:enz}
\end{figure}

A specific example of tracking and feedback control is shown in Fig. \ref{fig:enz}. Most importantly, when $ \left|2a^2t_0 N_s \ind \right| \leq 1$ we expect the ENZ response to be unique (see Sec. II of the supplemental material \cite{supplement}), and consequently we find that both feedback and tracking control yield identical solutions. Note that the derived driving field exhibits a high degree of complexity and bandwidth, which would be challenging to reproduce with current pulse shaping technology. There is a real prospect, however, of controlling the input electric field using the materials own response via feedback control.

\begin{figure}[ht]
\begin{center}
\includegraphics[width=1\columnwidth]{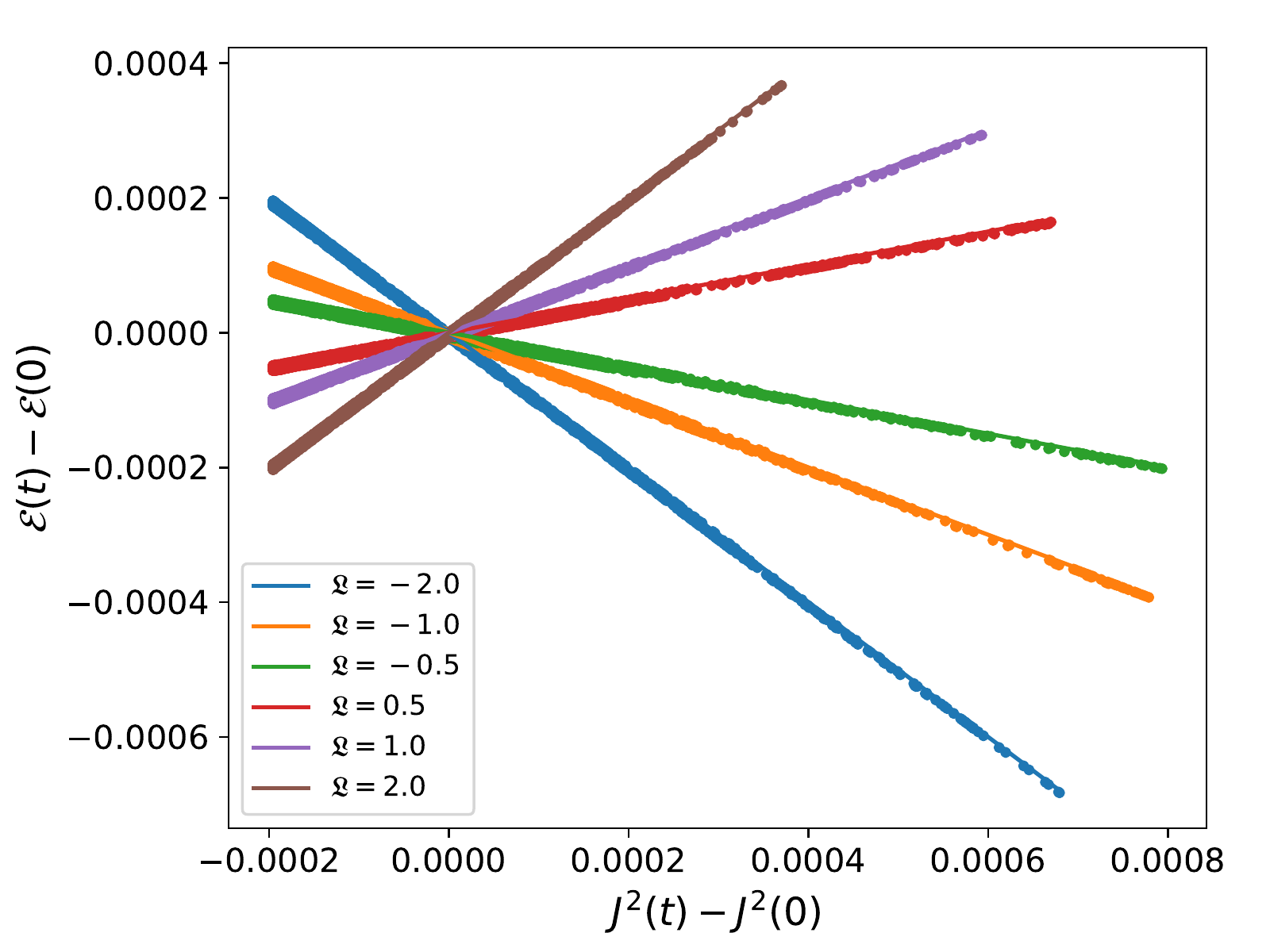}\end{center}
\caption{The relationship between the change in energy and the change in current induced by an ENZ pulse for a Fermi-Hubbard material simulated exactly with 10 sites, $\frac{U}{t_0} = 0.5$, $a = 4$ \r{A}. The slope of each line is $\frac{\ind}{2}$.}
\label{fig:energy-current}
\end{figure}

As predicted by Eq. \eqref{energycurrentrelationship} and demonstrated in Fig. \ref{fig:energy-current}, the relationship between the change in energy and the change in the square of the current is exactly linear and the slope increases with increasing $\ind$. The relationship holds for positive as well as negative values of $\ind$, and thus allows us to induce a non-Foster response by specifying $\ind < 0$.

\paragraph{Discussion.} In this letter, we have demonstrated the possibility of inducing an ENZ response in a large class of solid-state systems using either tracking or feedback control. Tracking control gives extremely precise control over the induced current, but faces significant hurdles to practical implementation due to the bandwidth of driving pulses. In contrast, feedback control replaces pulse shaping with amplification, and can therefore be implemented relatively straightforwardly. In the current work, we have assumed an instantaneous feedback, but this can easily be adapted to a finite optical path by preparing two identical copies of the system, and using feeding the response of one system to the initial pulse into the input of the second at a suitable time delay. Given the equivalence in the solutions they predict, it is possible that the most direct route to implementing many tracking control proposals will be to adapt them to a feedback procedure. It is also worth noting that the particular response one obtains from tracking will depend on the initial state, and hence the specific pump profile one applies before tracking. This offers a route to tailoring the response obtained. For example, one could generate a maximum or minimum bandwidth for the response by performing an optimisation over pump pulse parameters, conditioned on the desired property.

The equations developed here allow one to tune the scaling of the system response relative to the input field by a factor $\ind$. Analytically, this parameter is shown to be analogous to macroscopic inductance \eqref{energycurrentrelationship}, and we demonstrate that the predicted energy-current relationship is realized in simulations in Fig. \ref{fig:energy-current}. There are no restrictions on the sign of this inductance can take, meaning it is possible to specify $\ind < 0$ to induce a non-Foster reactance. Since coupling negative inductance materials to metamaterials has been shown to increase metamaterials' bandwidth, not only can we drive an ENZ response in these materials, but optical feedback control may offer an indirect route to realising ENZ behaviour in metamaterials by broadening their bandwidth.

The mapping between the ENZ response and an ideal inductor also hints at the possibility that feedback control may be employed to develop quantum optical analogs to other basic components of classical electrical circuits, such as resistance and capacitance. The possibility of an all optical computation has already been demonstrated in \cite{singleatom}, and with the limit of Moore's law being approached \cite{eeckhout2017moore}, the need for alternative computing platforms is pressing. Optical feedback control therefore represents a potential opportunity to take advantage of the inherent nonlinearities present in quantum optics, in order to develop computing at the nanoscale.

\emph{Acknowledgement.}
This work was supported by the W. M. Keck Foundation and by Army Research Office (ARO) (grant W911NF-19-1-0377, program manager Dr.~James Joseph, and cooperative agreement W911NF-21-2-0139). The views and conclusions contained in this document are those of the authors and should not be interpreted as representing the official policies, either expressed or implied, of ARO or the U.S. Government. The U.S. Government is authorized to reproduce and distribute reprints for Government purposes notwithstanding any copyright notation herein.

\bibliography{refs}

\begin{thebibliography}{91}%
\makeatletter
\providecommand \@ifxundefined [1]{%
 \@ifx{#1\undefined}
}%
\providecommand \@ifnum [1]{%
 \ifnum #1\expandafter \@firstoftwo
 \else \expandafter \@secondoftwo
 \fi
}%
\providecommand \@ifx [1]{%
 \ifx #1\expandafter \@firstoftwo
 \else \expandafter \@secondoftwo
 \fi
}%
\providecommand \natexlab [1]{#1}%
\providecommand \enquote  [1]{``#1''}%
\providecommand \bibnamefont  [1]{#1}%
\providecommand \bibfnamefont [1]{#1}%
\providecommand \citenamefont [1]{#1}%
\providecommand \href@noop [0]{\@secondoftwo}%
\providecommand \href [0]{\begingroup \@sanitize@url \@href}%
\providecommand \@href[1]{\@@startlink{#1}\@@href}%
\providecommand \@@href[1]{\endgroup#1\@@endlink}%
\providecommand \@sanitize@url [0]{\catcode `\\12\catcode `\$12\catcode
  `\&12\catcode `\#12\catcode `\^12\catcode `\_12\catcode `\%12\relax}%
\providecommand \@@startlink[1]{}%
\providecommand \@@endlink[0]{}%
\providecommand \url  [0]{\begingroup\@sanitize@url \@url }%
\providecommand \@url [1]{\endgroup\@href {#1}{\urlprefix }}%
\providecommand \urlprefix  [0]{URL }%
\providecommand \Eprint [0]{\href }%
\providecommand \doibase [0]{https://doi.org/}%
\providecommand \selectlanguage [0]{\@gobble}%
\providecommand \bibinfo  [0]{\@secondoftwo}%
\providecommand \bibfield  [0]{\@secondoftwo}%
\providecommand \translation [1]{[#1]}%
\providecommand \BibitemOpen [0]{}%
\providecommand \bibitemStop [0]{}%
\providecommand \bibitemNoStop [0]{.\EOS\space}%
\providecommand \EOS [0]{\spacefactor3000\relax}%
\providecommand \BibitemShut  [1]{\csname bibitem#1\endcsname}%
\let\auto@bib@innerbib\@empty
\bibitem [{\citenamefont {Wu}\ \emph {et~al.}(2021)\citenamefont {Wu},
  \citenamefont {Xie}, \citenamefont {Sha}, \citenamefont {Fu},\ and\
  \citenamefont {Li}}]{WuENZPhotonics}%
  \BibitemOpen
  \bibfield  {author} {\bibinfo {author} {\bibfnamefont {J.}~\bibnamefont
  {Wu}}, \bibinfo {author} {\bibfnamefont {Z.~T.}\ \bibnamefont {Xie}},
  \bibinfo {author} {\bibfnamefont {Y.}~\bibnamefont {Sha}}, \bibinfo {author}
  {\bibfnamefont {H.~Y.}\ \bibnamefont {Fu}},\ and\ \bibinfo {author}
  {\bibfnamefont {Q.}~\bibnamefont {Li}},\ }\bibfield  {title} {\bibinfo
  {title} {Epsilon-near-zero photonics: infinite potentials},\ }\href
  {https://doi.org/10.1364/PRJ.427246} {\bibfield  {journal} {\bibinfo
  {journal} {Photon. Res.}\ }\textbf {\bibinfo {volume} {9}},\ \bibinfo {pages}
  {1616} (\bibinfo {year} {2021})}\BibitemShut {NoStop}%
\bibitem [{\citenamefont {Ziolkowski}(2004)}]{ziolkowski2004propagation}%
  \BibitemOpen
  \bibfield  {author} {\bibinfo {author} {\bibfnamefont {R.~W.}\ \bibnamefont
  {Ziolkowski}},\ }\bibfield  {title} {\bibinfo {title} {Propagation in and
  scattering from a matched metamaterial having a zero index of refraction},\
  }\href@noop {} {\bibfield  {journal} {\bibinfo  {journal} {Physical Review
  E}\ }\textbf {\bibinfo {volume} {70}},\ \bibinfo {pages} {046608} (\bibinfo
  {year} {2004})}\BibitemShut {NoStop}%
\bibitem [{\citenamefont {Niu}\ \emph {et~al.}(2018)\citenamefont {Niu},
  \citenamefont {Hu}, \citenamefont {Chu},\ and\ \citenamefont
  {Gong}}]{ENZPhotonics}%
  \BibitemOpen
  \bibfield  {author} {\bibinfo {author} {\bibfnamefont {X.}~\bibnamefont
  {Niu}}, \bibinfo {author} {\bibfnamefont {X.}~\bibnamefont {Hu}}, \bibinfo
  {author} {\bibfnamefont {S.}~\bibnamefont {Chu}},\ and\ \bibinfo {author}
  {\bibfnamefont {Q.}~\bibnamefont {Gong}},\ }\bibfield  {title} {\bibinfo
  {title} {Epsilon-near-zero photonics: A new platform for integrated
  devices},\ }\href {https://doi.org/https://doi.org/10.1002/adom.201701292}
  {\bibfield  {journal} {\bibinfo  {journal} {Advanced Optical Materials}\
  }\textbf {\bibinfo {volume} {6}},\ \bibinfo {pages} {1701292} (\bibinfo
  {year} {2018})},\ \Eprint
  {https://arxiv.org/abs/https://onlinelibrary.wiley.com/doi/pdf/10.1002/adom.201701292}
  {https://onlinelibrary.wiley.com/doi/pdf/10.1002/adom.201701292} \BibitemShut
  {NoStop}%
\bibitem [{\citenamefont {Liberal}\ and\ \citenamefont
  {Engheta}(2017{\natexlab{a}})}]{nziPhotonics}%
  \BibitemOpen
  \bibfield  {author} {\bibinfo {author} {\bibfnamefont {I.}~\bibnamefont
  {Liberal}}\ and\ \bibinfo {author} {\bibfnamefont {N.}~\bibnamefont
  {Engheta}},\ }\bibfield  {title} {\bibinfo {title} {Near-zero refractive
  index photonics},\ }\href {https://doi.org/10.1038/nphoton.2017.13}
  {\bibfield  {journal} {\bibinfo  {journal} {Nature Photonics}\ }\textbf
  {\bibinfo {volume} {11}},\ \bibinfo {pages} {149} (\bibinfo {year}
  {2017}{\natexlab{a}})}\BibitemShut {NoStop}%
\bibitem [{\citenamefont {Liberal}\ and\ \citenamefont
  {Engheta}(2017{\natexlab{b}})}]{riseNZItech}%
  \BibitemOpen
  \bibfield  {author} {\bibinfo {author} {\bibfnamefont {I.}~\bibnamefont
  {Liberal}}\ and\ \bibinfo {author} {\bibfnamefont {N.}~\bibnamefont
  {Engheta}},\ }\bibfield  {title} {\bibinfo {title} {The rise of
  near-zero-index technologies},\ }\href
  {https://doi.org/10.1126/science.aaq0459} {\bibfield  {journal} {\bibinfo
  {journal} {Science}\ }\textbf {\bibinfo {volume} {358}},\ \bibinfo {pages}
  {1540} (\bibinfo {year} {2017}{\natexlab{b}})},\ \Eprint
  {https://arxiv.org/abs/https://www.science.org/doi/pdf/10.1126/science.aaq0459}
  {https://www.science.org/doi/pdf/10.1126/science.aaq0459} \BibitemShut
  {NoStop}%
\bibitem [{\citenamefont {Kinsey}\ \emph {et~al.}(2019)\citenamefont {Kinsey},
  \citenamefont {DeVault}, \citenamefont {Boltasseva},\ and\ \citenamefont
  {Shalaev}}]{nziPhotonicMaterials}%
  \BibitemOpen
  \bibfield  {author} {\bibinfo {author} {\bibfnamefont {N.}~\bibnamefont
  {Kinsey}}, \bibinfo {author} {\bibfnamefont {C.}~\bibnamefont {DeVault}},
  \bibinfo {author} {\bibfnamefont {A.}~\bibnamefont {Boltasseva}},\ and\
  \bibinfo {author} {\bibfnamefont {V.~M.}\ \bibnamefont {Shalaev}},\
  }\bibfield  {title} {\bibinfo {title} {Near-zero-index materials for
  photonics},\ }\href {https://doi.org/10.1038/s41578-019-0133-0} {\bibfield
  {journal} {\bibinfo  {journal} {Nature Reviews Materials}\ }\textbf {\bibinfo
  {volume} {4}},\ \bibinfo {pages} {742} (\bibinfo {year} {2019})}\BibitemShut
  {NoStop}%
\bibitem [{\citenamefont {Engheta}\ and\ \citenamefont
  {Ziolkowski}(2006)}]{engheta2006metamaterials}%
  \BibitemOpen
  \bibfield  {author} {\bibinfo {author} {\bibfnamefont {N.}~\bibnamefont
  {Engheta}}\ and\ \bibinfo {author} {\bibfnamefont {R.~W.}\ \bibnamefont
  {Ziolkowski}},\ }\href@noop {} {\emph {\bibinfo {title} {Metamaterials:
  physics and engineering explorations}}}\ (\bibinfo  {publisher} {John Wiley
  \& Sons},\ \bibinfo {year} {2006})\BibitemShut {NoStop}%
\bibitem [{\citenamefont {Engheta}(2013)}]{engheta2013pursuing}%
  \BibitemOpen
  \bibfield  {author} {\bibinfo {author} {\bibfnamefont {N.}~\bibnamefont
  {Engheta}},\ }\bibfield  {title} {\bibinfo {title} {Pursuing near-zero
  response},\ }\href@noop {} {\bibfield  {journal} {\bibinfo  {journal}
  {Science}\ }\textbf {\bibinfo {volume} {340}},\ \bibinfo {pages} {286}
  (\bibinfo {year} {2013})}\BibitemShut {NoStop}%
\bibitem [{\citenamefont {Vincenti}\ \emph {et~al.}(2020)\citenamefont
  {Vincenti}, \citenamefont {de~Ceglia},\ and\ \citenamefont
  {Scalora}}]{Vincenti:20}%
  \BibitemOpen
  \bibfield  {author} {\bibinfo {author} {\bibfnamefont {M.~A.}\ \bibnamefont
  {Vincenti}}, \bibinfo {author} {\bibfnamefont {D.}~\bibnamefont
  {de~Ceglia}},\ and\ \bibinfo {author} {\bibfnamefont {M.}~\bibnamefont
  {Scalora}},\ }\bibfield  {title} {\bibinfo {title} {{ENZ materials and
  anisotropy: enhancing nonlinear optical interactions at the nanoscale}},\
  }\href {https://doi.org/10.1364/OE.404107} {\bibfield  {journal} {\bibinfo
  {journal} {Opt. Express}\ }\textbf {\bibinfo {volume} {28}},\ \bibinfo
  {pages} {31180} (\bibinfo {year} {2020})}\BibitemShut {NoStop}%
\bibitem [{\citenamefont {Tong}\ \emph {et~al.}(2020)\citenamefont {Tong},
  \citenamefont {Ren}, \citenamefont {Tao},\ and\ \citenamefont
  {Tang}}]{Tong_2020}%
  \BibitemOpen
  \bibfield  {author} {\bibinfo {author} {\bibfnamefont {S.}~\bibnamefont
  {Tong}}, \bibinfo {author} {\bibfnamefont {C.}~\bibnamefont {Ren}}, \bibinfo
  {author} {\bibfnamefont {J.}~\bibnamefont {Tao}},\ and\ \bibinfo {author}
  {\bibfnamefont {W.}~\bibnamefont {Tang}},\ }\bibfield  {title} {\bibinfo
  {title} {Anisotropic index-near-zero metamaterials for enhanced directional
  acoustic emission},\ }\href {https://doi.org/10.1088/1361-6463/ab7df3}
  {\bibfield  {journal} {\bibinfo  {journal} {Journal of Physics D: Applied
  Physics}\ }\textbf {\bibinfo {volume} {53}},\ \bibinfo {pages} {265102}
  (\bibinfo {year} {2020})}\BibitemShut {NoStop}%
\bibitem [{\citenamefont {Vlasov}\ \emph {et~al.}(2001)\citenamefont {Vlasov},
  \citenamefont {Bo}, \citenamefont {Sturm},\ and\ \citenamefont
  {Norris}}]{vlasov2001silicon}%
  \BibitemOpen
  \bibfield  {author} {\bibinfo {author} {\bibfnamefont {Y.~A.}\ \bibnamefont
  {Vlasov}}, \bibinfo {author} {\bibfnamefont {X.-Z.}\ \bibnamefont {Bo}},
  \bibinfo {author} {\bibfnamefont {J.~C.}\ \bibnamefont {Sturm}},\ and\
  \bibinfo {author} {\bibfnamefont {D.~J.}\ \bibnamefont {Norris}},\ }\bibfield
   {title} {\bibinfo {title} {On-chip natural assembly of silicon photonic
  bandgap crystals},\ }\href {https://doi.org/10.1038/35104529} {\bibfield
  {journal} {\bibinfo  {journal} {Nature}\ }\textbf {\bibinfo {volume} {414}},\
  \bibinfo {pages} {289} (\bibinfo {year} {2001})}\BibitemShut {NoStop}%
\bibitem [{\citenamefont {Enoch}\ \emph {et~al.}(2002)\citenamefont {Enoch},
  \citenamefont {Tayeb}, \citenamefont {Sabouroux}, \citenamefont {Gu\'erin},\
  and\ \citenamefont {Vincent}}]{PhysRevLett.89.213902}%
  \BibitemOpen
  \bibfield  {author} {\bibinfo {author} {\bibfnamefont {S.}~\bibnamefont
  {Enoch}}, \bibinfo {author} {\bibfnamefont {G.}~\bibnamefont {Tayeb}},
  \bibinfo {author} {\bibfnamefont {P.}~\bibnamefont {Sabouroux}}, \bibinfo
  {author} {\bibfnamefont {N.}~\bibnamefont {Gu\'erin}},\ and\ \bibinfo
  {author} {\bibfnamefont {P.}~\bibnamefont {Vincent}},\ }\bibfield  {title}
  {\bibinfo {title} {A metamaterial for directive emission},\ }\href
  {https://doi.org/10.1103/PhysRevLett.89.213902} {\bibfield  {journal}
  {\bibinfo  {journal} {Phys. Rev. Lett.}\ }\textbf {\bibinfo {volume} {89}},\
  \bibinfo {pages} {213902} (\bibinfo {year} {2002})}\BibitemShut {NoStop}%
\bibitem [{\citenamefont {Guo}\ \emph {et~al.}(2017)\citenamefont {Guo},
  \citenamefont {Cui}, \citenamefont {Yao}, \citenamefont {Ye}, \citenamefont
  {Yang}, \citenamefont {Liu}, \citenamefont {Zhang}, \citenamefont {Liu},
  \citenamefont {Qiu},\ and\ \citenamefont
  {Hosono}}]{https://doi.org/10.1002/adma.201700754}%
  \BibitemOpen
  \bibfield  {author} {\bibinfo {author} {\bibfnamefont {Q.}~\bibnamefont
  {Guo}}, \bibinfo {author} {\bibfnamefont {Y.}~\bibnamefont {Cui}}, \bibinfo
  {author} {\bibfnamefont {Y.}~\bibnamefont {Yao}}, \bibinfo {author}
  {\bibfnamefont {Y.}~\bibnamefont {Ye}}, \bibinfo {author} {\bibfnamefont
  {Y.}~\bibnamefont {Yang}}, \bibinfo {author} {\bibfnamefont {X.}~\bibnamefont
  {Liu}}, \bibinfo {author} {\bibfnamefont {S.}~\bibnamefont {Zhang}}, \bibinfo
  {author} {\bibfnamefont {X.}~\bibnamefont {Liu}}, \bibinfo {author}
  {\bibfnamefont {J.}~\bibnamefont {Qiu}},\ and\ \bibinfo {author}
  {\bibfnamefont {H.}~\bibnamefont {Hosono}},\ }\bibfield  {title} {\bibinfo
  {title} {A solution-processed ultrafast optical switch based on a
  nanostructured epsilon-near-zero medium},\ }\href
  {https://doi.org/https://doi.org/10.1002/adma.201700754} {\bibfield
  {journal} {\bibinfo  {journal} {Advanced Materials}\ }\textbf {\bibinfo
  {volume} {29}},\ \bibinfo {pages} {1700754} (\bibinfo {year} {2017})},\
  \Eprint
  {https://arxiv.org/abs/https://onlinelibrary.wiley.com/doi/pdf/10.1002/adma.201700754}
  {https://onlinelibrary.wiley.com/doi/pdf/10.1002/adma.201700754} \BibitemShut
  {NoStop}%
\bibitem [{\citenamefont {Xie}\ \emph {et~al.}(2020)\citenamefont {Xie},
  \citenamefont {Wu}, \citenamefont {Fu},\ and\ \citenamefont {Li}}]{9144401}%
  \BibitemOpen
  \bibfield  {author} {\bibinfo {author} {\bibfnamefont {Z.~T.}\ \bibnamefont
  {Xie}}, \bibinfo {author} {\bibfnamefont {J.}~\bibnamefont {Wu}}, \bibinfo
  {author} {\bibfnamefont {H.~Y.}\ \bibnamefont {Fu}},\ and\ \bibinfo {author}
  {\bibfnamefont {Q.}~\bibnamefont {Li}},\ }\bibfield  {title} {\bibinfo
  {title} {Tunable electro- and all-optical switch based on epsilon-near-zero
  metasurface},\ }\href {https://doi.org/10.1109/JPHOT.2020.3010284} {\bibfield
   {journal} {\bibinfo  {journal} {IEEE Photonics Journal}\ }\textbf {\bibinfo
  {volume} {12}},\ \bibinfo {pages} {1} (\bibinfo {year} {2020})}\BibitemShut
  {NoStop}%
\bibitem [{\citenamefont {Bohn}\ \emph {et~al.}(2021)\citenamefont {Bohn},
  \citenamefont {Luk}, \citenamefont {Tollerton}, \citenamefont {Hutchings},
  \citenamefont {Brener}, \citenamefont {Horsley}, \citenamefont {Barnes},\
  and\ \citenamefont {Hendry}}]{bohn2021all}%
  \BibitemOpen
  \bibfield  {author} {\bibinfo {author} {\bibfnamefont {J.}~\bibnamefont
  {Bohn}}, \bibinfo {author} {\bibfnamefont {T.~S.}\ \bibnamefont {Luk}},
  \bibinfo {author} {\bibfnamefont {C.}~\bibnamefont {Tollerton}}, \bibinfo
  {author} {\bibfnamefont {S.~W.}\ \bibnamefont {Hutchings}}, \bibinfo {author}
  {\bibfnamefont {I.}~\bibnamefont {Brener}}, \bibinfo {author} {\bibfnamefont
  {S.}~\bibnamefont {Horsley}}, \bibinfo {author} {\bibfnamefont {W.~L.}\
  \bibnamefont {Barnes}},\ and\ \bibinfo {author} {\bibfnamefont
  {E.}~\bibnamefont {Hendry}},\ }\bibfield  {title} {\bibinfo {title}
  {All-optical switching of an epsilon-near-zero plasmon resonance in indium
  tin oxide},\ }\href {https://doi.org/10.1038/s41467-021-21332-y} {\bibfield
  {journal} {\bibinfo  {journal} {Nature Communications}\ }\textbf {\bibinfo
  {volume} {12}},\ \bibinfo {pages} {1017} (\bibinfo {year}
  {2021})}\BibitemShut {NoStop}%
\bibitem [{\citenamefont {Kuttruff}\ \emph {et~al.}(2020)\citenamefont
  {Kuttruff}, \citenamefont {Garoli}, \citenamefont {Allerbeck}, \citenamefont
  {Krahne}, \citenamefont {De~Luca}, \citenamefont {Brida}, \citenamefont
  {Caligiuri},\ and\ \citenamefont {Maccaferri}}]{kuttruff2020ultrafast}%
  \BibitemOpen
  \bibfield  {author} {\bibinfo {author} {\bibfnamefont {J.}~\bibnamefont
  {Kuttruff}}, \bibinfo {author} {\bibfnamefont {D.}~\bibnamefont {Garoli}},
  \bibinfo {author} {\bibfnamefont {J.}~\bibnamefont {Allerbeck}}, \bibinfo
  {author} {\bibfnamefont {R.}~\bibnamefont {Krahne}}, \bibinfo {author}
  {\bibfnamefont {A.}~\bibnamefont {De~Luca}}, \bibinfo {author} {\bibfnamefont
  {D.}~\bibnamefont {Brida}}, \bibinfo {author} {\bibfnamefont
  {V.}~\bibnamefont {Caligiuri}},\ and\ \bibinfo {author} {\bibfnamefont
  {N.}~\bibnamefont {Maccaferri}},\ }\bibfield  {title} {\bibinfo {title}
  {Ultrafast all-optical switching enabled by epsilon-near-zero-tailored
  absorption in metal-insulator nanocavities},\ }\href
  {https://doi.org/10.1038/s42005-020-0379-2} {\bibfield  {journal} {\bibinfo
  {journal} {Communications Physics}\ }\textbf {\bibinfo {volume} {3}},\
  \bibinfo {pages} {114} (\bibinfo {year} {2020})}\BibitemShut {NoStop}%
\bibitem [{\citenamefont {Zhang}\ \emph {et~al.}(2020)\citenamefont {Zhang},
  \citenamefont {Zu}, \citenamefont {Yang}, \citenamefont {Jiang},\ and\
  \citenamefont {Liu}}]{ZHANG2020106271}%
  \BibitemOpen
  \bibfield  {author} {\bibinfo {author} {\bibfnamefont {C.}~\bibnamefont
  {Zhang}}, \bibinfo {author} {\bibfnamefont {Y.}~\bibnamefont {Zu}}, \bibinfo
  {author} {\bibfnamefont {W.}~\bibnamefont {Yang}}, \bibinfo {author}
  {\bibfnamefont {S.}~\bibnamefont {Jiang}},\ and\ \bibinfo {author}
  {\bibfnamefont {J.}~\bibnamefont {Liu}},\ }\bibfield  {title} {\bibinfo
  {title} {Epsilon-near-zero medium for optical switches in ho solid-state
  laser at 2.06 μm},\ }\href
  {https://doi.org/https://doi.org/10.1016/j.optlastec.2020.106271} {\bibfield
  {journal} {\bibinfo  {journal} {Optics \& Laser Technology}\ }\textbf
  {\bibinfo {volume} {129}},\ \bibinfo {pages} {106271} (\bibinfo {year}
  {2020})}\BibitemShut {NoStop}%
\bibitem [{\citenamefont {Zhang}\ \emph {et~al.}(2019)\citenamefont {Zhang},
  \citenamefont {Liu}, \citenamefont {Hao},\ and\ \citenamefont
  {Liu}}]{Zhang_2019}%
  \BibitemOpen
  \bibfield  {author} {\bibinfo {author} {\bibfnamefont {Z.}~\bibnamefont
  {Zhang}}, \bibinfo {author} {\bibfnamefont {J.}~\bibnamefont {Liu}}, \bibinfo
  {author} {\bibfnamefont {Q.}~\bibnamefont {Hao}},\ and\ \bibinfo {author}
  {\bibfnamefont {J.}~\bibnamefont {Liu}},\ }\bibfield  {title} {\bibinfo
  {title} {Sensitive saturable absorber and optical switch of epsilon-near-zero
  medium},\ }\href {https://doi.org/10.7567/1882-0786/ab21e3} {\bibfield
  {journal} {\bibinfo  {journal} {Applied Physics Express}\ }\textbf {\bibinfo
  {volume} {12}},\ \bibinfo {pages} {065504} (\bibinfo {year}
  {2019})}\BibitemShut {NoStop}%
\bibitem [{\citenamefont {Lu}\ \emph {et~al.}(2012)\citenamefont {Lu},
  \citenamefont {Zhao},\ and\ \citenamefont {Shi}}]{6194256}%
  \BibitemOpen
  \bibfield  {author} {\bibinfo {author} {\bibfnamefont {Z.}~\bibnamefont
  {Lu}}, \bibinfo {author} {\bibfnamefont {W.}~\bibnamefont {Zhao}},\ and\
  \bibinfo {author} {\bibfnamefont {K.}~\bibnamefont {Shi}},\ }\bibfield
  {title} {\bibinfo {title} {Ultracompact electroabsorption modulators based on
  tunable epsilon-near-zero-slot waveguides},\ }\href
  {https://doi.org/10.1109/JPHOT.2012.2197742} {\bibfield  {journal} {\bibinfo
  {journal} {IEEE Photonics Journal}\ }\textbf {\bibinfo {volume} {4}},\
  \bibinfo {pages} {735} (\bibinfo {year} {2012})}\BibitemShut {NoStop}%
\bibitem [{\citenamefont {Zhao}\ \emph {et~al.}(2015)\citenamefont {Zhao},
  \citenamefont {Wang}, \citenamefont {Capretti}, \citenamefont {Negro},\ and\
  \citenamefont {Klamkin}}]{6967699}%
  \BibitemOpen
  \bibfield  {author} {\bibinfo {author} {\bibfnamefont {H.}~\bibnamefont
  {Zhao}}, \bibinfo {author} {\bibfnamefont {Y.}~\bibnamefont {Wang}}, \bibinfo
  {author} {\bibfnamefont {A.}~\bibnamefont {Capretti}}, \bibinfo {author}
  {\bibfnamefont {L.~D.}\ \bibnamefont {Negro}},\ and\ \bibinfo {author}
  {\bibfnamefont {J.}~\bibnamefont {Klamkin}},\ }\bibfield  {title} {\bibinfo
  {title} {Broadband electroabsorption modulators design based on
  epsilon-near-zero indium tin oxide},\ }\href
  {https://doi.org/10.1109/JSTQE.2014.2375153} {\bibfield  {journal} {\bibinfo
  {journal} {IEEE Journal of Selected Topics in Quantum Electronics}\ }\textbf
  {\bibinfo {volume} {21}},\ \bibinfo {pages} {192} (\bibinfo {year}
  {2015})}\BibitemShut {NoStop}%
\bibitem [{\citenamefont {Koch}\ \emph {et~al.}(2016)\citenamefont {Koch},
  \citenamefont {Hoessbacher}, \citenamefont {Niegemann}, \citenamefont
  {Hafner},\ and\ \citenamefont {Leuthold}}]{7386561}%
  \BibitemOpen
  \bibfield  {author} {\bibinfo {author} {\bibfnamefont {U.}~\bibnamefont
  {Koch}}, \bibinfo {author} {\bibfnamefont {C.}~\bibnamefont {Hoessbacher}},
  \bibinfo {author} {\bibfnamefont {J.}~\bibnamefont {Niegemann}}, \bibinfo
  {author} {\bibfnamefont {C.}~\bibnamefont {Hafner}},\ and\ \bibinfo {author}
  {\bibfnamefont {J.}~\bibnamefont {Leuthold}},\ }\bibfield  {title} {\bibinfo
  {title} {Digital plasmonic absorption modulator exploiting epsilon-near-zero
  in transparent conducting oxides},\ }\href
  {https://doi.org/10.1109/JPHOT.2016.2518861} {\bibfield  {journal} {\bibinfo
  {journal} {IEEE Photonics Journal}\ }\textbf {\bibinfo {volume} {8}},\
  \bibinfo {pages} {1} (\bibinfo {year} {2016})}\BibitemShut {NoStop}%
\bibitem [{\citenamefont {Baek}\ \emph {et~al.}(2015)\citenamefont {Baek},
  \citenamefont {You},\ and\ \citenamefont {Yu}}]{Baek:15}%
  \BibitemOpen
  \bibfield  {author} {\bibinfo {author} {\bibfnamefont {J.}~\bibnamefont
  {Baek}}, \bibinfo {author} {\bibfnamefont {J.-B.}\ \bibnamefont {You}},\ and\
  \bibinfo {author} {\bibfnamefont {K.}~\bibnamefont {Yu}},\ }\bibfield
  {title} {\bibinfo {title} {Free-carrier electro-refraction modulation based
  on a silicon slot waveguide with ito},\ }\href
  {https://doi.org/10.1364/OE.23.015863} {\bibfield  {journal} {\bibinfo
  {journal} {Opt. Express}\ }\textbf {\bibinfo {volume} {23}},\ \bibinfo
  {pages} {15863} (\bibinfo {year} {2015})}\BibitemShut {NoStop}%
\bibitem [{\citenamefont {Lee}\ \emph {et~al.}(2014)\citenamefont {Lee},
  \citenamefont {Papadakis}, \citenamefont {Burgos}, \citenamefont {Chander},
  \citenamefont {Kriesch}, \citenamefont {Pala}, \citenamefont {Peschel},\ and\
  \citenamefont {Atwater}}]{lee2014nanoscale}%
  \BibitemOpen
  \bibfield  {author} {\bibinfo {author} {\bibfnamefont {H.~W.}\ \bibnamefont
  {Lee}}, \bibinfo {author} {\bibfnamefont {G.}~\bibnamefont {Papadakis}},
  \bibinfo {author} {\bibfnamefont {S.~P.}\ \bibnamefont {Burgos}}, \bibinfo
  {author} {\bibfnamefont {K.}~\bibnamefont {Chander}}, \bibinfo {author}
  {\bibfnamefont {A.}~\bibnamefont {Kriesch}}, \bibinfo {author} {\bibfnamefont
  {R.}~\bibnamefont {Pala}}, \bibinfo {author} {\bibfnamefont {U.}~\bibnamefont
  {Peschel}},\ and\ \bibinfo {author} {\bibfnamefont {H.~A.}\ \bibnamefont
  {Atwater}},\ }\bibfield  {title} {\bibinfo {title} {Nanoscale conducting
  oxide plasmostor},\ }\href {https://doi.org/10.1021/nl502998z} {\bibfield
  {journal} {\bibinfo  {journal} {Nano Letters}\ }\textbf {\bibinfo {volume}
  {14}},\ \bibinfo {pages} {6463} (\bibinfo {year} {2014})}\BibitemShut
  {NoStop}%
\bibitem [{\citenamefont {Vasudev}\ \emph {et~al.}(2013)\citenamefont
  {Vasudev}, \citenamefont {Kang}, \citenamefont {Park}, \citenamefont {Liu},\
  and\ \citenamefont {Brongersma}}]{Vasudev:13}%
  \BibitemOpen
  \bibfield  {author} {\bibinfo {author} {\bibfnamefont {A.~P.}\ \bibnamefont
  {Vasudev}}, \bibinfo {author} {\bibfnamefont {J.-H.}\ \bibnamefont {Kang}},
  \bibinfo {author} {\bibfnamefont {J.}~\bibnamefont {Park}}, \bibinfo {author}
  {\bibfnamefont {X.}~\bibnamefont {Liu}},\ and\ \bibinfo {author}
  {\bibfnamefont {M.~L.}\ \bibnamefont {Brongersma}},\ }\bibfield  {title}
  {\bibinfo {title} {Electro-optical modulation of a silicon waveguide with an
  ``epsilon-near-zero'' material},\ }\href
  {https://doi.org/10.1364/OE.21.026387} {\bibfield  {journal} {\bibinfo
  {journal} {Opt. Express}\ }\textbf {\bibinfo {volume} {21}},\ \bibinfo
  {pages} {26387} (\bibinfo {year} {2013})}\BibitemShut {NoStop}%
\bibitem [{\citenamefont {Liu}\ \emph {et~al.}(2018)\citenamefont {Liu},
  \citenamefont {Zang}, \citenamefont {Kang}, \citenamefont {Park},
  \citenamefont {Harris}, \citenamefont {Kik},\ and\ \citenamefont
  {Brongersma}}]{liu2018acs}%
  \BibitemOpen
  \bibfield  {author} {\bibinfo {author} {\bibfnamefont {X.}~\bibnamefont
  {Liu}}, \bibinfo {author} {\bibfnamefont {K.}~\bibnamefont {Zang}}, \bibinfo
  {author} {\bibfnamefont {J.-H.}\ \bibnamefont {Kang}}, \bibinfo {author}
  {\bibfnamefont {J.}~\bibnamefont {Park}}, \bibinfo {author} {\bibfnamefont
  {J.~S.}\ \bibnamefont {Harris}}, \bibinfo {author} {\bibfnamefont {P.~G.}\
  \bibnamefont {Kik}},\ and\ \bibinfo {author} {\bibfnamefont {M.~L.}\
  \bibnamefont {Brongersma}},\ }\bibfield  {title} {\bibinfo {title}
  {Epsilon-near-zero si slot-waveguide modulator},\ }\href
  {https://doi.org/10.1021/acsphotonics.8b00945} {\bibfield  {journal}
  {\bibinfo  {journal} {ACS Photonics}\ }\textbf {\bibinfo {volume} {5}},\
  \bibinfo {pages} {4484} (\bibinfo {year} {2018})}\BibitemShut {NoStop}%
\bibitem [{\citenamefont {Rizza}\ \emph {et~al.}(2011)\citenamefont {Rizza},
  \citenamefont {Di~Falco},\ and\ \citenamefont
  {Ciattoni}}]{doi:10.1063/1.3665414}%
  \BibitemOpen
  \bibfield  {author} {\bibinfo {author} {\bibfnamefont {C.}~\bibnamefont
  {Rizza}}, \bibinfo {author} {\bibfnamefont {A.}~\bibnamefont {Di~Falco}},\
  and\ \bibinfo {author} {\bibfnamefont {A.}~\bibnamefont {Ciattoni}},\
  }\bibfield  {title} {\bibinfo {title} {Gain assisted nanocomposite
  multilayers with near zero permittivity modulus at visible frequencies},\
  }\href {https://doi.org/10.1063/1.3665414} {\bibfield  {journal} {\bibinfo
  {journal} {Applied Physics Letters}\ }\textbf {\bibinfo {volume} {99}},\
  \bibinfo {pages} {221107} (\bibinfo {year} {2011})},\ \Eprint
  {https://arxiv.org/abs/https://doi.org/10.1063/1.3665414}
  {https://doi.org/10.1063/1.3665414} \BibitemShut {NoStop}%
\bibitem [{\citenamefont {Maas}\ \emph {et~al.}(2013)\citenamefont {Maas},
  \citenamefont {Parsons}, \citenamefont {Engheta},\ and\ \citenamefont
  {Polman}}]{maasVisibleENZ}%
  \BibitemOpen
  \bibfield  {author} {\bibinfo {author} {\bibfnamefont {R.}~\bibnamefont
  {Maas}}, \bibinfo {author} {\bibfnamefont {J.}~\bibnamefont {Parsons}},
  \bibinfo {author} {\bibfnamefont {N.}~\bibnamefont {Engheta}},\ and\ \bibinfo
  {author} {\bibfnamefont {A.}~\bibnamefont {Polman}},\ }\bibfield  {title}
  {\bibinfo {title} {Experimental realization of an epsilon-near-zero
  metamaterial at visible wavelengths},\ }\href
  {https://doi.org/10.1038/nphoton.2013.256} {\bibfield  {journal} {\bibinfo
  {journal} {Nature Photonics}\ }\textbf {\bibinfo {volume} {7}},\ \bibinfo
  {pages} {907} (\bibinfo {year} {2013})}\BibitemShut {NoStop}%
\bibitem [{\citenamefont {Zhao}\ and\ \citenamefont
  {Xie}(2019)}]{zhao2019novel}%
  \BibitemOpen
  \bibfield  {author} {\bibinfo {author} {\bibfnamefont {L.}~\bibnamefont
  {Zhao}}\ and\ \bibinfo {author} {\bibfnamefont {H.}~\bibnamefont {Xie}},\
  }\bibfield  {title} {\bibinfo {title} {A novel optical \textepsilon
  -near-zero material realized by multi-layered ag/sic film structures},\
  }\href {https://doi.org/https://doi.org/10.1016/j.ijleo.2019.02.059}
  {\bibfield  {journal} {\bibinfo  {journal} {Optik}\ }\textbf {\bibinfo
  {volume} {183}},\ \bibinfo {pages} {513} (\bibinfo {year}
  {2019})}\BibitemShut {NoStop}%
\bibitem [{\citenamefont {Kelley}\ \emph {et~al.}(2019)\citenamefont {Kelley},
  \citenamefont {Runnerstrom}, \citenamefont {Sachet}, \citenamefont {Shelton},
  \citenamefont {Grimley}, \citenamefont {Klump}, \citenamefont {LeBeau},
  \citenamefont {Sitar}, \citenamefont {Suen}, \citenamefont {Padilla},\ and\
  \citenamefont {Maria}}]{kelley2019Multiple}%
  \BibitemOpen
  \bibfield  {author} {\bibinfo {author} {\bibfnamefont {K.~P.}\ \bibnamefont
  {Kelley}}, \bibinfo {author} {\bibfnamefont {E.~L.}\ \bibnamefont
  {Runnerstrom}}, \bibinfo {author} {\bibfnamefont {E.}~\bibnamefont {Sachet}},
  \bibinfo {author} {\bibfnamefont {C.~T.}\ \bibnamefont {Shelton}}, \bibinfo
  {author} {\bibfnamefont {E.~D.}\ \bibnamefont {Grimley}}, \bibinfo {author}
  {\bibfnamefont {A.}~\bibnamefont {Klump}}, \bibinfo {author} {\bibfnamefont
  {J.~M.}\ \bibnamefont {LeBeau}}, \bibinfo {author} {\bibfnamefont
  {Z.}~\bibnamefont {Sitar}}, \bibinfo {author} {\bibfnamefont {J.~Y.}\
  \bibnamefont {Suen}}, \bibinfo {author} {\bibfnamefont {W.~J.}\ \bibnamefont
  {Padilla}},\ and\ \bibinfo {author} {\bibfnamefont {J.-P.}\ \bibnamefont
  {Maria}},\ }\bibfield  {title} {\bibinfo {title} {Multiple epsilon-near-zero
  resonances in multilayered cadmium oxide: Designing metamaterial-like optical
  properties in monolithic materials},\ }\href
  {https://doi.org/10.1021/acsphotonics.9b00367} {\bibfield  {journal}
  {\bibinfo  {journal} {ACS Photonics}\ }\textbf {\bibinfo {volume} {6}},\
  \bibinfo {pages} {1139} (\bibinfo {year} {2019})}\BibitemShut {NoStop}%
\bibitem [{\citenamefont {Dai}\ and\ \citenamefont {Zhang}(2015)}]{Dai:15}%
  \BibitemOpen
  \bibfield  {author} {\bibinfo {author} {\bibfnamefont {D.}~\bibnamefont
  {Dai}}\ and\ \bibinfo {author} {\bibfnamefont {M.}~\bibnamefont {Zhang}},\
  }\bibfield  {title} {\bibinfo {title} {Mode hybridization and conversion in
  silicon-on-insulator nanowires with angled sidewalls},\ }\href
  {https://doi.org/10.1364/OE.23.032452} {\bibfield  {journal} {\bibinfo
  {journal} {Opt. Express}\ }\textbf {\bibinfo {volume} {23}},\ \bibinfo
  {pages} {32452} (\bibinfo {year} {2015})}\BibitemShut {NoStop}%
\bibitem [{\citenamefont {Caligiuri}\ \emph {et~al.}(2019)\citenamefont
  {Caligiuri}, \citenamefont {Palei}, \citenamefont {Biffi},\ and\
  \citenamefont {Krahne}}]{CaligiuriPaleiBiffiKrahne}%
  \BibitemOpen
  \bibfield  {author} {\bibinfo {author} {\bibfnamefont {V.}~\bibnamefont
  {Caligiuri}}, \bibinfo {author} {\bibfnamefont {M.}~\bibnamefont {Palei}},
  \bibinfo {author} {\bibfnamefont {G.}~\bibnamefont {Biffi}},\ and\ \bibinfo
  {author} {\bibfnamefont {R.}~\bibnamefont {Krahne}},\ }\href
  {https://doi.org/doi:10.1515/nanoph-2019-0054} {\bibfield  {journal}
  {\bibinfo  {journal} {Nanophotonics}\ }\textbf {\bibinfo {volume} {8}},\
  \bibinfo {pages} {1505} (\bibinfo {year} {2019})}\BibitemShut {NoStop}%
\bibitem [{\citenamefont {Rashed}\ \emph {et~al.}(2020)\citenamefont {Rashed},
  \citenamefont {Yildiz}, \citenamefont {Ayyagari},\ and\ \citenamefont
  {Caglayan}}]{PhysRevB.101.165301}%
  \BibitemOpen
  \bibfield  {author} {\bibinfo {author} {\bibfnamefont {A.~R.}\ \bibnamefont
  {Rashed}}, \bibinfo {author} {\bibfnamefont {B.~C.}\ \bibnamefont {Yildiz}},
  \bibinfo {author} {\bibfnamefont {S.~R.}\ \bibnamefont {Ayyagari}},\ and\
  \bibinfo {author} {\bibfnamefont {H.}~\bibnamefont {Caglayan}},\ }\bibfield
  {title} {\bibinfo {title} {Hot electron dynamics in ultrafast multilayer
  epsilon-near-zero metamaterials},\ }\href
  {https://doi.org/10.1103/PhysRevB.101.165301} {\bibfield  {journal} {\bibinfo
   {journal} {Phys. Rev. B}\ }\textbf {\bibinfo {volume} {101}},\ \bibinfo
  {pages} {165301} (\bibinfo {year} {2020})}\BibitemShut {NoStop}%
\bibitem [{\citenamefont {Koivurova}\ \emph {et~al.}(2020)\citenamefont
  {Koivurova}, \citenamefont {Hakala}, \citenamefont {Turunen}, \citenamefont
  {Friberg}, \citenamefont {Ornigotti},\ and\ \citenamefont
  {Caglayan}}]{Koivurova_2020}%
  \BibitemOpen
  \bibfield  {author} {\bibinfo {author} {\bibfnamefont {M.}~\bibnamefont
  {Koivurova}}, \bibinfo {author} {\bibfnamefont {T.}~\bibnamefont {Hakala}},
  \bibinfo {author} {\bibfnamefont {J.}~\bibnamefont {Turunen}}, \bibinfo
  {author} {\bibfnamefont {A.~T.}\ \bibnamefont {Friberg}}, \bibinfo {author}
  {\bibfnamefont {M.}~\bibnamefont {Ornigotti}},\ and\ \bibinfo {author}
  {\bibfnamefont {H.}~\bibnamefont {Caglayan}},\ }\bibfield  {title} {\bibinfo
  {title} {Metamaterials designed for enhanced enz properties},\ }\href
  {https://doi.org/10.1088/1367-2630/abb387} {\bibfield  {journal} {\bibinfo
  {journal} {New Journal of Physics}\ }\textbf {\bibinfo {volume} {22}},\
  \bibinfo {pages} {093054} (\bibinfo {year} {2020})}\BibitemShut {NoStop}%
\bibitem [{\citenamefont {Lee}\ and\ \citenamefont
  {Kee}(2021)}]{LEE2021412598}%
  \BibitemOpen
  \bibfield  {author} {\bibinfo {author} {\bibfnamefont {Y.~G.}\ \bibnamefont
  {Lee}}\ and\ \bibinfo {author} {\bibfnamefont {C.-S.}\ \bibnamefont {Kee}},\
  }\bibfield  {title} {\bibinfo {title} {Constant cutoff frequency of a
  two-dimensional photonic crystal composed of metallic rods and
  epsilon-near-zero materials},\ }\href
  {https://doi.org/https://doi.org/10.1016/j.physb.2020.412598} {\bibfield
  {journal} {\bibinfo  {journal} {Physica B: Condensed Matter}\ }\textbf
  {\bibinfo {volume} {600}},\ \bibinfo {pages} {412598} (\bibinfo {year}
  {2021})}\BibitemShut {NoStop}%
\bibitem [{\citenamefont {Anderegg}\ \emph {et~al.}(1971)\citenamefont
  {Anderegg}, \citenamefont {Feuerbacher},\ and\ \citenamefont
  {Fitton}}]{anderegg1971optically}%
  \BibitemOpen
  \bibfield  {author} {\bibinfo {author} {\bibfnamefont {M.}~\bibnamefont
  {Anderegg}}, \bibinfo {author} {\bibfnamefont {B.}~\bibnamefont
  {Feuerbacher}},\ and\ \bibinfo {author} {\bibfnamefont {B.}~\bibnamefont
  {Fitton}},\ }\bibfield  {title} {\bibinfo {title} {Optically excited
  longitudinal plasmons in potassium},\ }\href@noop {} {\bibfield  {journal}
  {\bibinfo  {journal} {Physical Review Letters}\ }\textbf {\bibinfo {volume}
  {27}},\ \bibinfo {pages} {1565} (\bibinfo {year} {1971})}\BibitemShut
  {NoStop}%
\bibitem [{\citenamefont {Spitzer}\ \emph {et~al.}(1959)\citenamefont
  {Spitzer}, \citenamefont {Kleinman},\ and\ \citenamefont
  {Walsh}}]{spitzer1959infrared}%
  \BibitemOpen
  \bibfield  {author} {\bibinfo {author} {\bibfnamefont {W.}~\bibnamefont
  {Spitzer}}, \bibinfo {author} {\bibfnamefont {D.}~\bibnamefont {Kleinman}},\
  and\ \bibinfo {author} {\bibfnamefont {D.}~\bibnamefont {Walsh}},\ }\bibfield
   {title} {\bibinfo {title} {Infrared properties of hexagonal silicon
  carbide},\ }\href@noop {} {\bibfield  {journal} {\bibinfo  {journal}
  {Physical Review}\ }\textbf {\bibinfo {volume} {113}},\ \bibinfo {pages}
  {127} (\bibinfo {year} {1959})}\BibitemShut {NoStop}%
\bibitem [{\citenamefont {Korobkin}\ \emph {et~al.}(2006)\citenamefont
  {Korobkin}, \citenamefont {Urzhumov},\ and\ \citenamefont
  {Shvets}}]{korobkin2006enhanced}%
  \BibitemOpen
  \bibfield  {author} {\bibinfo {author} {\bibfnamefont {D.}~\bibnamefont
  {Korobkin}}, \bibinfo {author} {\bibfnamefont {Y.}~\bibnamefont {Urzhumov}},\
  and\ \bibinfo {author} {\bibfnamefont {G.}~\bibnamefont {Shvets}},\
  }\bibfield  {title} {\bibinfo {title} {Enhanced near-field resolution in
  midinfrared using metamaterials},\ }\href@noop {} {\bibfield  {journal}
  {\bibinfo  {journal} {JOSA B}\ }\textbf {\bibinfo {volume} {23}},\ \bibinfo
  {pages} {468} (\bibinfo {year} {2006})}\BibitemShut {NoStop}%
\bibitem [{\citenamefont {Caldwell}\ \emph {et~al.}(2015)\citenamefont
  {Caldwell}, \citenamefont {Lindsay}, \citenamefont {Giannini}, \citenamefont
  {Vurgaftman}, \citenamefont {Reinecke}, \citenamefont {Maier},\ and\
  \citenamefont {Glembocki}}]{caldwell2015low}%
  \BibitemOpen
  \bibfield  {author} {\bibinfo {author} {\bibfnamefont {J.~D.}\ \bibnamefont
  {Caldwell}}, \bibinfo {author} {\bibfnamefont {L.}~\bibnamefont {Lindsay}},
  \bibinfo {author} {\bibfnamefont {V.}~\bibnamefont {Giannini}}, \bibinfo
  {author} {\bibfnamefont {I.}~\bibnamefont {Vurgaftman}}, \bibinfo {author}
  {\bibfnamefont {T.~L.}\ \bibnamefont {Reinecke}}, \bibinfo {author}
  {\bibfnamefont {S.~A.}\ \bibnamefont {Maier}},\ and\ \bibinfo {author}
  {\bibfnamefont {O.~J.}\ \bibnamefont {Glembocki}},\ }\bibfield  {title}
  {\bibinfo {title} {Low-loss, infrared and terahertz nanophotonics using
  surface phonon polaritons},\ }\href@noop {} {\bibfield  {journal} {\bibinfo
  {journal} {Nanophotonics}\ }\textbf {\bibinfo {volume} {4}},\ \bibinfo
  {pages} {44} (\bibinfo {year} {2015})}\BibitemShut {NoStop}%
\bibitem [{\citenamefont {Kim}\ \emph {et~al.}(2016)\citenamefont {Kim},
  \citenamefont {Dutta}, \citenamefont {Naik}, \citenamefont {Giles},
  \citenamefont {Bezares}, \citenamefont {Ellis}, \citenamefont {Tischler},
  \citenamefont {Mahmoud}, \citenamefont {Caglayan}, \citenamefont {Glembocki}
  \emph {et~al.}}]{kim2016role}%
  \BibitemOpen
  \bibfield  {author} {\bibinfo {author} {\bibfnamefont {J.}~\bibnamefont
  {Kim}}, \bibinfo {author} {\bibfnamefont {A.}~\bibnamefont {Dutta}}, \bibinfo
  {author} {\bibfnamefont {G.~V.}\ \bibnamefont {Naik}}, \bibinfo {author}
  {\bibfnamefont {A.~J.}\ \bibnamefont {Giles}}, \bibinfo {author}
  {\bibfnamefont {F.~J.}\ \bibnamefont {Bezares}}, \bibinfo {author}
  {\bibfnamefont {C.~T.}\ \bibnamefont {Ellis}}, \bibinfo {author}
  {\bibfnamefont {J.~G.}\ \bibnamefont {Tischler}}, \bibinfo {author}
  {\bibfnamefont {A.~M.}\ \bibnamefont {Mahmoud}}, \bibinfo {author}
  {\bibfnamefont {H.}~\bibnamefont {Caglayan}}, \bibinfo {author}
  {\bibfnamefont {O.~J.}\ \bibnamefont {Glembocki}}, \emph {et~al.},\
  }\bibfield  {title} {\bibinfo {title} {Role of epsilon-near-zero substrates
  in the optical response of plasmonic antennas},\ }\href@noop {} {\bibfield
  {journal} {\bibinfo  {journal} {Optica}\ }\textbf {\bibinfo {volume} {3}},\
  \bibinfo {pages} {339} (\bibinfo {year} {2016})}\BibitemShut {NoStop}%
\bibitem [{\citenamefont {Naik}\ \emph {et~al.}(2011)\citenamefont {Naik},
  \citenamefont {Kim},\ and\ \citenamefont {Boltasseva}}]{naik2011oxides}%
  \BibitemOpen
  \bibfield  {author} {\bibinfo {author} {\bibfnamefont {G.~V.}\ \bibnamefont
  {Naik}}, \bibinfo {author} {\bibfnamefont {J.}~\bibnamefont {Kim}},\ and\
  \bibinfo {author} {\bibfnamefont {A.}~\bibnamefont {Boltasseva}},\ }\bibfield
   {title} {\bibinfo {title} {Oxides and nitrides as alternative plasmonic
  materials in the optical range},\ }\href@noop {} {\bibfield  {journal}
  {\bibinfo  {journal} {Optical materials express}\ }\textbf {\bibinfo {volume}
  {1}},\ \bibinfo {pages} {1090} (\bibinfo {year} {2011})}\BibitemShut
  {NoStop}%
\bibitem [{\citenamefont {Naik}\ \emph {et~al.}(2013)\citenamefont {Naik},
  \citenamefont {Shalaev},\ and\ \citenamefont
  {Boltasseva}}]{naik2013alternative}%
  \BibitemOpen
  \bibfield  {author} {\bibinfo {author} {\bibfnamefont {G.~V.}\ \bibnamefont
  {Naik}}, \bibinfo {author} {\bibfnamefont {V.~M.}\ \bibnamefont {Shalaev}},\
  and\ \bibinfo {author} {\bibfnamefont {A.}~\bibnamefont {Boltasseva}},\
  }\bibfield  {title} {\bibinfo {title} {Alternative plasmonic materials:
  beyond gold and silver},\ }\href@noop {} {\bibfield  {journal} {\bibinfo
  {journal} {Advanced Materials}\ }\textbf {\bibinfo {volume} {25}},\ \bibinfo
  {pages} {3264} (\bibinfo {year} {2013})}\BibitemShut {NoStop}%
\bibitem [{\citenamefont {Kinsey}\ \emph {et~al.}(2015)\citenamefont {Kinsey},
  \citenamefont {DeVault}, \citenamefont {Kim}, \citenamefont {Ferrera},
  \citenamefont {Shalaev},\ and\ \citenamefont
  {Boltasseva}}]{kinsey2015epsilon}%
  \BibitemOpen
  \bibfield  {author} {\bibinfo {author} {\bibfnamefont {N.}~\bibnamefont
  {Kinsey}}, \bibinfo {author} {\bibfnamefont {C.}~\bibnamefont {DeVault}},
  \bibinfo {author} {\bibfnamefont {J.}~\bibnamefont {Kim}}, \bibinfo {author}
  {\bibfnamefont {M.}~\bibnamefont {Ferrera}}, \bibinfo {author} {\bibfnamefont
  {V.}~\bibnamefont {Shalaev}},\ and\ \bibinfo {author} {\bibfnamefont
  {A.}~\bibnamefont {Boltasseva}},\ }\bibfield  {title} {\bibinfo {title}
  {Epsilon-near-zero al-doped zno for ultrafast switching at telecom
  wavelengths},\ }\href@noop {} {\bibfield  {journal} {\bibinfo  {journal}
  {Optica}\ }\textbf {\bibinfo {volume} {2}},\ \bibinfo {pages} {616} (\bibinfo
  {year} {2015})}\BibitemShut {NoStop}%
\bibitem [{\citenamefont {Ou}\ \emph {et~al.}(2014)\citenamefont {Ou},
  \citenamefont {So}, \citenamefont {Adamo}, \citenamefont {Sulaev},
  \citenamefont {Wang},\ and\ \citenamefont {Zheludev}}]{ou2014ultraviolet}%
  \BibitemOpen
  \bibfield  {author} {\bibinfo {author} {\bibfnamefont {J.-Y.}\ \bibnamefont
  {Ou}}, \bibinfo {author} {\bibfnamefont {J.-K.}\ \bibnamefont {So}}, \bibinfo
  {author} {\bibfnamefont {G.}~\bibnamefont {Adamo}}, \bibinfo {author}
  {\bibfnamefont {A.}~\bibnamefont {Sulaev}}, \bibinfo {author} {\bibfnamefont
  {L.}~\bibnamefont {Wang}},\ and\ \bibinfo {author} {\bibfnamefont {N.~I.}\
  \bibnamefont {Zheludev}},\ }\bibfield  {title} {\bibinfo {title} {Ultraviolet
  and visible range plasmonics in the topological insulator bi1. 5sb0. 5te1.
  8se1. 2},\ }\href@noop {} {\bibfield  {journal} {\bibinfo  {journal} {Nature
  communications}\ }\textbf {\bibinfo {volume} {5}},\ \bibinfo {pages} {1}
  (\bibinfo {year} {2014})}\BibitemShut {NoStop}%
\bibitem [{\citenamefont {Javani}\ and\ \citenamefont
  {Stockman}(2016)}]{PhysRevLett.117.107404}%
  \BibitemOpen
  \bibfield  {author} {\bibinfo {author} {\bibfnamefont {M.~H.}\ \bibnamefont
  {Javani}}\ and\ \bibinfo {author} {\bibfnamefont {M.~I.}\ \bibnamefont
  {Stockman}},\ }\bibfield  {title} {\bibinfo {title} {Real and imaginary
  properties of epsilon-near-zero materials},\ }\href
  {https://doi.org/10.1103/PhysRevLett.117.107404} {\bibfield  {journal}
  {\bibinfo  {journal} {Phys. Rev. Lett.}\ }\textbf {\bibinfo {volume} {117}},\
  \bibinfo {pages} {107404} (\bibinfo {year} {2016})}\BibitemShut {NoStop}%
\bibitem [{\citenamefont {Avignon-Meseldzija}\ \emph
  {et~al.}(2017)\citenamefont {Avignon-Meseldzija}, \citenamefont {Lepetit},
  \citenamefont {Ferreira},\ and\ \citenamefont {Boust}}]{avignon2017negative}%
  \BibitemOpen
  \bibfield  {author} {\bibinfo {author} {\bibfnamefont {E.}~\bibnamefont
  {Avignon-Meseldzija}}, \bibinfo {author} {\bibfnamefont {T.}~\bibnamefont
  {Lepetit}}, \bibinfo {author} {\bibfnamefont {P.~M.}\ \bibnamefont
  {Ferreira}},\ and\ \bibinfo {author} {\bibfnamefont {F.}~\bibnamefont
  {Boust}},\ }\bibfield  {title} {\bibinfo {title} {Negative inductance
  circuits for metamaterial bandwidth enhancement},\ }\href@noop {} {\bibfield
  {journal} {\bibinfo  {journal} {EPJ Applied Metamaterials}\ }\textbf
  {\bibinfo {volume} {4}},\ \bibinfo {pages} {11} (\bibinfo {year}
  {2017})}\BibitemShut {NoStop}%
\bibitem [{\citenamefont {Youla}\ \emph {et~al.}(1959)\citenamefont {Youla},
  \citenamefont {Castriota},\ and\ \citenamefont {Carlin}}]{youla1959bounded}%
  \BibitemOpen
  \bibfield  {author} {\bibinfo {author} {\bibfnamefont {D.}~\bibnamefont
  {Youla}}, \bibinfo {author} {\bibfnamefont {L.}~\bibnamefont {Castriota}},\
  and\ \bibinfo {author} {\bibfnamefont {H.}~\bibnamefont {Carlin}},\
  }\bibfield  {title} {\bibinfo {title} {Bounded real scattering matrices and
  the foundations of linear passive network theory},\ }\href@noop {} {\bibfield
   {journal} {\bibinfo  {journal} {IRE Transactions on Circuit Theory}\
  }\textbf {\bibinfo {volume} {6}},\ \bibinfo {pages} {102} (\bibinfo {year}
  {1959})}\BibitemShut {NoStop}%
\bibitem [{\citenamefont {Tretyakov}(2001)}]{tretyakov2001meta}%
  \BibitemOpen
  \bibfield  {author} {\bibinfo {author} {\bibfnamefont {S.}~\bibnamefont
  {Tretyakov}},\ }\bibfield  {title} {\bibinfo {title} {Meta-materials with
  wideband negative permittivity and permeability},\ }\href@noop {} {\bibfield
  {journal} {\bibinfo  {journal} {Microwave and Optical Technology Letters}\
  }\textbf {\bibinfo {volume} {31}},\ \bibinfo {pages} {163} (\bibinfo {year}
  {2001})}\BibitemShut {NoStop}%
\bibitem [{\citenamefont {Tretyakov}\ and\ \citenamefont
  {Maslovski}(2007)}]{4231250}%
  \BibitemOpen
  \bibfield  {author} {\bibinfo {author} {\bibfnamefont {S.~A.}\ \bibnamefont
  {Tretyakov}}\ and\ \bibinfo {author} {\bibfnamefont {S.~I.}\ \bibnamefont
  {Maslovski}},\ }\bibfield  {title} {\bibinfo {title} {Veselago materials:
  What is possible and impossible about the dispersion of the constitutive
  parameters},\ }\href {https://doi.org/10.1109/MAP.2007.370980} {\bibfield
  {journal} {\bibinfo  {journal} {IEEE Antennas and Propagation Magazine}\
  }\textbf {\bibinfo {volume} {49}},\ \bibinfo {pages} {37} (\bibinfo {year}
  {2007})}\BibitemShut {NoStop}%
\bibitem [{\citenamefont {Montgomery}\ \emph {et~al.}(1987)\citenamefont
  {Montgomery}, \citenamefont {Dicke}, \citenamefont {Purcell},\ and\
  \citenamefont {Purcell}}]{montgomery1987principles}%
  \BibitemOpen
  \bibfield  {author} {\bibinfo {author} {\bibfnamefont {C.~G.}\ \bibnamefont
  {Montgomery}}, \bibinfo {author} {\bibfnamefont {R.~H.}\ \bibnamefont
  {Dicke}}, \bibinfo {author} {\bibfnamefont {E.~M.}\ \bibnamefont {Purcell}},\
  and\ \bibinfo {author} {\bibfnamefont {E.~M.}\ \bibnamefont {Purcell}},\
  }\href@noop {} {\emph {\bibinfo {title} {Principles of microwave
  circuits}}},\ \bibinfo {number} {25}\ (\bibinfo  {publisher} {Iet},\ \bibinfo
  {year} {1987})\BibitemShut {NoStop}%
\bibitem [{\citenamefont {Hrabar}\ \emph {et~al.}(2011)\citenamefont {Hrabar},
  \citenamefont {Krois}, \citenamefont {Bonic},\ and\ \citenamefont
  {Kiricenko}}]{hrabarnegativecapacitor}%
  \BibitemOpen
  \bibfield  {author} {\bibinfo {author} {\bibfnamefont {S.}~\bibnamefont
  {Hrabar}}, \bibinfo {author} {\bibfnamefont {I.}~\bibnamefont {Krois}},
  \bibinfo {author} {\bibfnamefont {I.}~\bibnamefont {Bonic}},\ and\ \bibinfo
  {author} {\bibfnamefont {A.}~\bibnamefont {Kiricenko}},\ }\bibfield  {title}
  {\bibinfo {title} {Negative capacitor paves the way to ultra-broadband
  metamaterials},\ }\href {https://doi.org/10.1063/1.3671366} {\bibfield
  {journal} {\bibinfo  {journal} {Applied Physics Letters}\ }\textbf {\bibinfo
  {volume} {99}},\ \bibinfo {pages} {254103} (\bibinfo {year} {2011})},\
  \Eprint {https://arxiv.org/abs/https://doi.org/10.1063/1.3671366}
  {https://doi.org/10.1063/1.3671366} \BibitemShut {NoStop}%
\bibitem [{\citenamefont {Hrabar}\ \emph {et~al.}(2010)\citenamefont {Hrabar},
  \citenamefont {Krois},\ and\ \citenamefont {Kiricenko}}]{hrabar2010towards}%
  \BibitemOpen
  \bibfield  {author} {\bibinfo {author} {\bibfnamefont {S.}~\bibnamefont
  {Hrabar}}, \bibinfo {author} {\bibfnamefont {I.}~\bibnamefont {Krois}},\ and\
  \bibinfo {author} {\bibfnamefont {A.}~\bibnamefont {Kiricenko}},\ }\bibfield
  {title} {\bibinfo {title} {Towards active dispersionless enz metamaterial for
  cloaking applications},\ }\href@noop {} {\bibfield  {journal} {\bibinfo
  {journal} {Metamaterials}\ }\textbf {\bibinfo {volume} {4}},\ \bibinfo
  {pages} {89} (\bibinfo {year} {2010})}\BibitemShut {NoStop}%
\bibitem [{\citenamefont {d'Alessandro}(2021)}]{d2021introduction}%
  \BibitemOpen
  \bibfield  {author} {\bibinfo {author} {\bibfnamefont {D.}~\bibnamefont
  {d'Alessandro}},\ }\href@noop {} {\emph {\bibinfo {title} {Introduction to
  quantum control and dynamics}}}\ (\bibinfo  {publisher} {Chapman and
  hall/CRC},\ \bibinfo {year} {2021})\BibitemShut {NoStop}%
\bibitem [{\citenamefont {Dong}\ and\ \citenamefont
  {Petersen}(2010)}]{dong2010quantum}%
  \BibitemOpen
  \bibfield  {author} {\bibinfo {author} {\bibfnamefont {D.}~\bibnamefont
  {Dong}}\ and\ \bibinfo {author} {\bibfnamefont {I.~R.}\ \bibnamefont
  {Petersen}},\ }\bibfield  {title} {\bibinfo {title} {Quantum control theory
  and applications: a survey},\ }\href@noop {} {\bibfield  {journal} {\bibinfo
  {journal} {IET control theory \& applications}\ }\textbf {\bibinfo {volume}
  {4}},\ \bibinfo {pages} {2651} (\bibinfo {year} {2010})}\BibitemShut
  {NoStop}%
\bibitem [{\citenamefont {Rothman}\ \emph {et~al.}(2005)\citenamefont
  {Rothman}, \citenamefont {Ho},\ and\ \citenamefont
  {Rabitz}}]{PhysRevA.72.023416}%
  \BibitemOpen
  \bibfield  {author} {\bibinfo {author} {\bibfnamefont {A.}~\bibnamefont
  {Rothman}}, \bibinfo {author} {\bibfnamefont {T.-S.}\ \bibnamefont {Ho}},\
  and\ \bibinfo {author} {\bibfnamefont {H.}~\bibnamefont {Rabitz}},\
  }\bibfield  {title} {\bibinfo {title} {Observable-preserving control of
  quantum dynamics over a family of related systems},\ }\href
  {https://doi.org/10.1103/PhysRevA.72.023416} {\bibfield  {journal} {\bibinfo
  {journal} {Phys. Rev. A}\ }\textbf {\bibinfo {volume} {72}},\ \bibinfo
  {pages} {023416} (\bibinfo {year} {2005})}\BibitemShut {NoStop}%
\bibitem [{\citenamefont {Magann}\ \emph {et~al.}(2018)\citenamefont {Magann},
  \citenamefont {Ho},\ and\ \citenamefont {Rabitz}}]{PhysRevA.98.043429}%
  \BibitemOpen
  \bibfield  {author} {\bibinfo {author} {\bibfnamefont {A.}~\bibnamefont
  {Magann}}, \bibinfo {author} {\bibfnamefont {T.-S.}\ \bibnamefont {Ho}},\
  and\ \bibinfo {author} {\bibfnamefont {H.}~\bibnamefont {Rabitz}},\
  }\bibfield  {title} {\bibinfo {title} {Singularity-free quantum tracking
  control of molecular rotor orientation},\ }\href
  {https://doi.org/10.1103/PhysRevA.98.043429} {\bibfield  {journal} {\bibinfo
  {journal} {Phys. Rev. A}\ }\textbf {\bibinfo {volume} {98}},\ \bibinfo
  {pages} {043429} (\bibinfo {year} {2018})}\BibitemShut {NoStop}%
\bibitem [{\citenamefont {Caneva}\ \emph {et~al.}(2011)\citenamefont {Caneva},
  \citenamefont {Calarco},\ and\ \citenamefont
  {Montangero}}]{PhysRevA.84.022326}%
  \BibitemOpen
  \bibfield  {author} {\bibinfo {author} {\bibfnamefont {T.}~\bibnamefont
  {Caneva}}, \bibinfo {author} {\bibfnamefont {T.}~\bibnamefont {Calarco}},\
  and\ \bibinfo {author} {\bibfnamefont {S.}~\bibnamefont {Montangero}},\
  }\bibfield  {title} {\bibinfo {title} {Chopped random-basis quantum
  optimization},\ }\href {https://doi.org/10.1103/PhysRevA.84.022326}
  {\bibfield  {journal} {\bibinfo  {journal} {Phys. Rev. A}\ }\textbf {\bibinfo
  {volume} {84}},\ \bibinfo {pages} {022326} (\bibinfo {year}
  {2011})}\BibitemShut {NoStop}%
\bibitem [{\citenamefont {Campos}\ \emph {et~al.}(2017)\citenamefont {Campos},
  \citenamefont {Bondar}, \citenamefont {Cabrera},\ and\ \citenamefont
  {Rabitz}}]{PhysRevLett.118.083201}%
  \BibitemOpen
  \bibfield  {author} {\bibinfo {author} {\bibfnamefont {A.~G.}\ \bibnamefont
  {Campos}}, \bibinfo {author} {\bibfnamefont {D.~I.}\ \bibnamefont {Bondar}},
  \bibinfo {author} {\bibfnamefont {R.}~\bibnamefont {Cabrera}},\ and\ \bibinfo
  {author} {\bibfnamefont {H.~A.}\ \bibnamefont {Rabitz}},\ }\bibfield  {title}
  {\bibinfo {title} {How to make distinct dynamical systems appear spectrally
  identical},\ }\href {https://doi.org/10.1103/PhysRevLett.118.083201}
  {\bibfield  {journal} {\bibinfo  {journal} {Phys. Rev. Lett.}\ }\textbf
  {\bibinfo {volume} {118}},\ \bibinfo {pages} {083201} (\bibinfo {year}
  {2017})}\BibitemShut {NoStop}%
\bibitem [{\citenamefont {Zhu}\ and\ \citenamefont
  {Rabitz}(2003)}]{doi:10.1063/1.1582847}%
  \BibitemOpen
  \bibfield  {author} {\bibinfo {author} {\bibfnamefont {W.}~\bibnamefont
  {Zhu}}\ and\ \bibinfo {author} {\bibfnamefont {H.}~\bibnamefont {Rabitz}},\
  }\bibfield  {title} {\bibinfo {title} {Quantum control design via adaptive
  tracking},\ }\href {https://doi.org/10.1063/1.1582847} {\bibfield  {journal}
  {\bibinfo  {journal} {The Journal of Chemical Physics}\ }\textbf {\bibinfo
  {volume} {119}},\ \bibinfo {pages} {3619} (\bibinfo {year} {2003})},\ \Eprint
  {https://arxiv.org/abs/https://doi.org/10.1063/1.1582847}
  {https://doi.org/10.1063/1.1582847} \BibitemShut {NoStop}%
\bibitem [{\citenamefont {Zhu}\ \emph {et~al.}(1999)\citenamefont {Zhu},
  \citenamefont {Smit},\ and\ \citenamefont {Rabitz}}]{doi:10.1063/1.477857}%
  \BibitemOpen
  \bibfield  {author} {\bibinfo {author} {\bibfnamefont {W.}~\bibnamefont
  {Zhu}}, \bibinfo {author} {\bibfnamefont {M.}~\bibnamefont {Smit}},\ and\
  \bibinfo {author} {\bibfnamefont {H.}~\bibnamefont {Rabitz}},\ }\bibfield
  {title} {\bibinfo {title} {Managing singular behavior in the tracking control
  of quantum dynamical observables},\ }\href {https://doi.org/10.1063/1.477857}
  {\bibfield  {journal} {\bibinfo  {journal} {The Journal of Chemical Physics}\
  }\textbf {\bibinfo {volume} {110}},\ \bibinfo {pages} {1905} (\bibinfo {year}
  {1999})},\ \Eprint {https://arxiv.org/abs/https://doi.org/10.1063/1.477857}
  {https://doi.org/10.1063/1.477857} \BibitemShut {NoStop}%
\bibitem [{\citenamefont {McCaul}\ \emph
  {et~al.}(2020{\natexlab{a}})\citenamefont {McCaul}, \citenamefont
  {Orthodoxou}, \citenamefont {Jacobs}, \citenamefont {Booth},\ and\
  \citenamefont {Bondar}}]{Gerard1}%
  \BibitemOpen
  \bibfield  {author} {\bibinfo {author} {\bibfnamefont {G.}~\bibnamefont
  {McCaul}}, \bibinfo {author} {\bibfnamefont {C.}~\bibnamefont {Orthodoxou}},
  \bibinfo {author} {\bibfnamefont {K.}~\bibnamefont {Jacobs}}, \bibinfo
  {author} {\bibfnamefont {G.~H.}\ \bibnamefont {Booth}},\ and\ \bibinfo
  {author} {\bibfnamefont {D.~I.}\ \bibnamefont {Bondar}},\ }\bibfield  {title}
  {\bibinfo {title} {Driven imposters: Controlling expectations in many-body
  systems},\ }\href {https://doi.org/10.1103/PhysRevLett.124.183201} {\bibfield
   {journal} {\bibinfo  {journal} {Phys. Rev. Lett.}\ }\textbf {\bibinfo
  {volume} {124}},\ \bibinfo {pages} {183201} (\bibinfo {year}
  {2020}{\natexlab{a}})}\BibitemShut {NoStop}%
\bibitem [{\citenamefont {McCaul}\ \emph
  {et~al.}(2020{\natexlab{b}})\citenamefont {McCaul}, \citenamefont
  {Orthodoxou}, \citenamefont {Jacobs}, \citenamefont {Booth},\ and\
  \citenamefont {Bondar}}]{Gerard2}%
  \BibitemOpen
  \bibfield  {author} {\bibinfo {author} {\bibfnamefont {G.}~\bibnamefont
  {McCaul}}, \bibinfo {author} {\bibfnamefont {C.}~\bibnamefont {Orthodoxou}},
  \bibinfo {author} {\bibfnamefont {K.}~\bibnamefont {Jacobs}}, \bibinfo
  {author} {\bibfnamefont {G.~H.}\ \bibnamefont {Booth}},\ and\ \bibinfo
  {author} {\bibfnamefont {D.~I.}\ \bibnamefont {Bondar}},\ }\bibfield  {title}
  {\bibinfo {title} {Controlling arbitrary observables in correlated many-body
  systems},\ }\href {https://doi.org/10.1103/PhysRevA.101.053408} {\bibfield
  {journal} {\bibinfo  {journal} {Phys. Rev. A}\ }\textbf {\bibinfo {volume}
  {101}},\ \bibinfo {pages} {053408} (\bibinfo {year}
  {2020}{\natexlab{b}})}\BibitemShut {NoStop}%
\bibitem [{\citenamefont {Magann}\ \emph
  {et~al.}(2022{\natexlab{a}})\citenamefont {Magann}, \citenamefont {McCaul},
  \citenamefont {Rabitz},\ and\ \citenamefont {Bondar}}]{Magann2022}%
  \BibitemOpen
  \bibfield  {author} {\bibinfo {author} {\bibfnamefont {A.~B.}\ \bibnamefont
  {Magann}}, \bibinfo {author} {\bibfnamefont {G.}~\bibnamefont {McCaul}},
  \bibinfo {author} {\bibfnamefont {H.~A.}\ \bibnamefont {Rabitz}},\ and\
  \bibinfo {author} {\bibfnamefont {D.~I.}\ \bibnamefont {Bondar}},\ }\bibfield
   {title} {\bibinfo {title} {Sequential optical response suppression for
  chemical mixture characterization},\ }\href
  {https://doi.org/10.22331/q-2022-01-20-626} {\bibfield  {journal} {\bibinfo
  {journal} {Quantum}\ }\textbf {\bibinfo {volume} {6}},\ \bibinfo {pages}
  {626} (\bibinfo {year} {2022}{\natexlab{a}})}\BibitemShut {NoStop}%
\bibitem [{\citenamefont {Magann}\ \emph {et~al.}(2023)\citenamefont {Magann},
  \citenamefont {Ho}, \citenamefont {Arenz},\ and\ \citenamefont
  {Rabitz}}]{Magann2023}%
  \BibitemOpen
  \bibfield  {author} {\bibinfo {author} {\bibfnamefont {A.~B.}\ \bibnamefont
  {Magann}}, \bibinfo {author} {\bibfnamefont {T.-S.}\ \bibnamefont {Ho}},
  \bibinfo {author} {\bibfnamefont {C.}~\bibnamefont {Arenz}},\ and\ \bibinfo
  {author} {\bibfnamefont {H.~A.}\ \bibnamefont {Rabitz}},\ }\href@noop {}
  {\bibinfo {title} {Quantum tracking control of the orientation of symmetric
  top molecules}} (\bibinfo {year} {2023}),\ \Eprint
  {https://arxiv.org/abs/arXiv:2301.04255} {arXiv:2301.04255} \BibitemShut
  {NoStop}%
\bibitem [{\citenamefont {McCaul}\ \emph {et~al.}(2021)\citenamefont {McCaul},
  \citenamefont {King},\ and\ \citenamefont {Bondar}}]{twinning_fields}%
  \BibitemOpen
  \bibfield  {author} {\bibinfo {author} {\bibfnamefont {G.}~\bibnamefont
  {McCaul}}, \bibinfo {author} {\bibfnamefont {A.~F.}\ \bibnamefont {King}},\
  and\ \bibinfo {author} {\bibfnamefont {D.~I.}\ \bibnamefont {Bondar}},\
  }\bibfield  {title} {\bibinfo {title} {Optical indistinguishability via
  twinning fields},\ }\href {https://doi.org/10.1103/PhysRevLett.127.113201}
  {\bibfield  {journal} {\bibinfo  {journal} {Phys. Rev. Lett.}\ }\textbf
  {\bibinfo {volume} {127}},\ \bibinfo {pages} {113201} (\bibinfo {year}
  {2021})}\BibitemShut {NoStop}%
\bibitem [{\citenamefont {McCaul}\ \emph {et~al.}(2022)\citenamefont {McCaul},
  \citenamefont {King},\ and\ \citenamefont {Bondar}}]{nonuniqueness}%
  \BibitemOpen
  \bibfield  {author} {\bibinfo {author} {\bibfnamefont {G.}~\bibnamefont
  {McCaul}}, \bibinfo {author} {\bibfnamefont {A.~F.}\ \bibnamefont {King}},\
  and\ \bibinfo {author} {\bibfnamefont {D.~I.}\ \bibnamefont {Bondar}},\
  }\bibfield  {title} {\bibinfo {title} {Non-uniqueness of driving fields
  generating non-linear optical response},\ }\href
  {https://doi.org/https://doi.org/10.1002/andp.202100523} {\bibfield
  {journal} {\bibinfo  {journal} {Annalen der Physik}\ }\textbf {\bibinfo
  {volume} {534}},\ \bibinfo {pages} {2100523} (\bibinfo {year}
  {2022})}\BibitemShut {NoStop}%
\bibitem [{\citenamefont {{\AA}str{\"o}m}\ and\ \citenamefont
  {Murray}(2021)}]{aastrom2021feedback}%
  \BibitemOpen
  \bibfield  {author} {\bibinfo {author} {\bibfnamefont {K.~J.}\ \bibnamefont
  {{\AA}str{\"o}m}}\ and\ \bibinfo {author} {\bibfnamefont {R.~M.}\
  \bibnamefont {Murray}},\ }\href@noop {} {\emph {\bibinfo {title} {Feedback
  systems: an introduction for scientists and engineers}}}\ (\bibinfo
  {publisher} {Princeton university press},\ \bibinfo {year}
  {2021})\BibitemShut {NoStop}%
\bibitem [{\citenamefont {Powers}\ and\ \citenamefont
  {Nicastri}(2000)}]{powers2000automotive}%
  \BibitemOpen
  \bibfield  {author} {\bibinfo {author} {\bibfnamefont {W.~F.}\ \bibnamefont
  {Powers}}\ and\ \bibinfo {author} {\bibfnamefont {P.~R.}\ \bibnamefont
  {Nicastri}},\ }\bibfield  {title} {\bibinfo {title} {Automotive vehicle
  control challenges in the 21st century},\ }\href@noop {} {\bibfield
  {journal} {\bibinfo  {journal} {Control engineering practice}\ }\textbf
  {\bibinfo {volume} {8}},\ \bibinfo {pages} {605} (\bibinfo {year}
  {2000})}\BibitemShut {NoStop}%
\bibitem [{\citenamefont {Kiencke}\ and\ \citenamefont
  {Nielsen}(2000)}]{kiencke2000automotive}%
  \BibitemOpen
  \bibfield  {author} {\bibinfo {author} {\bibfnamefont {U.}~\bibnamefont
  {Kiencke}}\ and\ \bibinfo {author} {\bibfnamefont {L.}~\bibnamefont
  {Nielsen}},\ }\href@noop {} {\bibinfo {title} {Automotive control systems:
  for engine, driveline, and vehicle}} (\bibinfo {year} {2000})\BibitemShut
  {NoStop}%
\bibitem [{\citenamefont {Barron}\ and\ \citenamefont
  {Powers}(1996)}]{barron1996role}%
  \BibitemOpen
  \bibfield  {author} {\bibinfo {author} {\bibfnamefont {M.~B.}\ \bibnamefont
  {Barron}}\ and\ \bibinfo {author} {\bibfnamefont {W.~F.}\ \bibnamefont
  {Powers}},\ }\bibfield  {title} {\bibinfo {title} {The role of electronic
  controls for future automotive mechatronic systems},\ }\href@noop {}
  {\bibfield  {journal} {\bibinfo  {journal} {IEEE/ASME Transactions on
  mechatronics}\ }\textbf {\bibinfo {volume} {1}},\ \bibinfo {pages} {80}
  (\bibinfo {year} {1996})}\BibitemShut {NoStop}%
\bibitem [{\citenamefont {Low}\ \emph {et~al.}(2002)\citenamefont {Low},
  \citenamefont {Paganini},\ and\ \citenamefont {Doyle}}]{low2002internet}%
  \BibitemOpen
  \bibfield  {author} {\bibinfo {author} {\bibfnamefont {S.~H.}\ \bibnamefont
  {Low}}, \bibinfo {author} {\bibfnamefont {F.}~\bibnamefont {Paganini}},\ and\
  \bibinfo {author} {\bibfnamefont {J.~C.}\ \bibnamefont {Doyle}},\ }\bibfield
  {title} {\bibinfo {title} {Internet congestion control},\ }\href@noop {}
  {\bibfield  {journal} {\bibinfo  {journal} {IEEE control systems magazine}\
  }\textbf {\bibinfo {volume} {22}},\ \bibinfo {pages} {28} (\bibinfo {year}
  {2002})}\BibitemShut {NoStop}%
\bibitem [{\citenamefont {Tanenbaum}(1981)}]{tanenbaum1981network}%
  \BibitemOpen
  \bibfield  {author} {\bibinfo {author} {\bibfnamefont {A.~S.}\ \bibnamefont
  {Tanenbaum}},\ }\bibfield  {title} {\bibinfo {title} {Network protocols},\
  }\href {https://doi.org/10.1145/356859.356864} {\bibfield  {journal}
  {\bibinfo  {journal} {ACM Comput. Surv.}\ }\textbf {\bibinfo {volume} {13}},\
  \bibinfo {pages} {453–489} (\bibinfo {year} {1981})}\BibitemShut {NoStop}%
\bibitem [{\citenamefont {Jacobson}(1988)}]{jacobson1988congestion}%
  \BibitemOpen
  \bibfield  {author} {\bibinfo {author} {\bibfnamefont {V.}~\bibnamefont
  {Jacobson}},\ }\bibfield  {title} {\bibinfo {title} {Congestion avoidance and
  control},\ }\href@noop {} {\bibfield  {journal} {\bibinfo  {journal} {ACM
  SIGCOMM computer communication review}\ }\textbf {\bibinfo {volume} {18}},\
  \bibinfo {pages} {314} (\bibinfo {year} {1988})}\BibitemShut {NoStop}%
\bibitem [{\citenamefont {Hellerstein}\ \emph {et~al.}(2004)\citenamefont
  {Hellerstein}, \citenamefont {Diao}, \citenamefont {Parekh},\ and\
  \citenamefont {Tilbury}}]{hellerstein2004feedback}%
  \BibitemOpen
  \bibfield  {author} {\bibinfo {author} {\bibfnamefont {J.~L.}\ \bibnamefont
  {Hellerstein}}, \bibinfo {author} {\bibfnamefont {Y.}~\bibnamefont {Diao}},
  \bibinfo {author} {\bibfnamefont {S.}~\bibnamefont {Parekh}},\ and\ \bibinfo
  {author} {\bibfnamefont {D.~M.}\ \bibnamefont {Tilbury}},\ }\href@noop {}
  {\emph {\bibinfo {title} {Feedback control of computing systems}}}\ (\bibinfo
   {publisher} {John Wiley \& Sons},\ \bibinfo {year} {2004})\BibitemShut
  {NoStop}%
\bibitem [{\citenamefont {Sarid}(1991)}]{sarid1991atomic}%
  \BibitemOpen
  \bibfield  {author} {\bibinfo {author} {\bibfnamefont {D.}~\bibnamefont
  {Sarid}},\ }\href@noop {} {\bibinfo {title} {Atomic force microscopy}}
  (\bibinfo {year} {1991})\BibitemShut {NoStop}%
\bibitem [{\citenamefont {Schitter}\ \emph {et~al.}(2001)\citenamefont
  {Schitter}, \citenamefont {Menold}, \citenamefont {Knapp}, \citenamefont
  {Allg{\"o}wer},\ and\ \citenamefont {Stemmer}}]{schitter2001high}%
  \BibitemOpen
  \bibfield  {author} {\bibinfo {author} {\bibfnamefont {G.}~\bibnamefont
  {Schitter}}, \bibinfo {author} {\bibfnamefont {P.}~\bibnamefont {Menold}},
  \bibinfo {author} {\bibfnamefont {H.}~\bibnamefont {Knapp}}, \bibinfo
  {author} {\bibfnamefont {F.}~\bibnamefont {Allg{\"o}wer}},\ and\ \bibinfo
  {author} {\bibfnamefont {A.}~\bibnamefont {Stemmer}},\ }\bibfield  {title}
  {\bibinfo {title} {High performance feedback for fast scanning atomic force
  microscopes},\ }\href@noop {} {\bibfield  {journal} {\bibinfo  {journal}
  {Review of Scientific Instruments}\ }\textbf {\bibinfo {volume} {72}},\
  \bibinfo {pages} {3320} (\bibinfo {year} {2001})}\BibitemShut {NoStop}%
\bibitem [{\citenamefont {Magann}\ \emph
  {et~al.}(2022{\natexlab{b}})\citenamefont {Magann}, \citenamefont {Rudinger},
  \citenamefont {Grace},\ and\ \citenamefont {Sarovar}}]{PhysRevA.106.062414}%
  \BibitemOpen
  \bibfield  {author} {\bibinfo {author} {\bibfnamefont {A.~B.}\ \bibnamefont
  {Magann}}, \bibinfo {author} {\bibfnamefont {K.~M.}\ \bibnamefont
  {Rudinger}}, \bibinfo {author} {\bibfnamefont {M.~D.}\ \bibnamefont
  {Grace}},\ and\ \bibinfo {author} {\bibfnamefont {M.}~\bibnamefont
  {Sarovar}},\ }\bibfield  {title} {\bibinfo {title} {Lyapunov-control-inspired
  strategies for quantum combinatorial optimization},\ }\href
  {https://doi.org/10.1103/PhysRevA.106.062414} {\bibfield  {journal} {\bibinfo
   {journal} {Phys. Rev. A}\ }\textbf {\bibinfo {volume} {106}},\ \bibinfo
  {pages} {062414} (\bibinfo {year} {2022}{\natexlab{b}})}\BibitemShut
  {NoStop}%
\bibitem [{\citenamefont {Magann}\ \emph
  {et~al.}(2022{\natexlab{c}})\citenamefont {Magann}, \citenamefont {Rudinger},
  \citenamefont {Grace},\ and\ \citenamefont
  {Sarovar}}]{PhysRevLett.129.250502}%
  \BibitemOpen
  \bibfield  {author} {\bibinfo {author} {\bibfnamefont {A.~B.}\ \bibnamefont
  {Magann}}, \bibinfo {author} {\bibfnamefont {K.~M.}\ \bibnamefont
  {Rudinger}}, \bibinfo {author} {\bibfnamefont {M.~D.}\ \bibnamefont
  {Grace}},\ and\ \bibinfo {author} {\bibfnamefont {M.}~\bibnamefont
  {Sarovar}},\ }\bibfield  {title} {\bibinfo {title} {Feedback-based quantum
  optimization},\ }\href {https://doi.org/10.1103/PhysRevLett.129.250502}
  {\bibfield  {journal} {\bibinfo  {journal} {Phys. Rev. Lett.}\ }\textbf
  {\bibinfo {volume} {129}},\ \bibinfo {pages} {250502} (\bibinfo {year}
  {2022}{\natexlab{c}})}\BibitemShut {NoStop}%
\bibitem [{\citenamefont {Peierls}(1933)}]{Peierls1933}%
  \BibitemOpen
  \bibfield  {author} {\bibinfo {author} {\bibfnamefont {R.}~\bibnamefont
  {Peierls}},\ }\bibfield  {title} {\bibinfo {title} {Zur theorie des
  diamagnetismus von leitungselektronen},\ }\href
  {https://doi.org/10.1007/bf01342591} {\bibfield  {journal} {\bibinfo
  {journal} {Zeitschrift f{\"u}r Physik}\ }\textbf {\bibinfo {volume} {80}},\
  \bibinfo {pages} {763} (\bibinfo {year} {1933})}\BibitemShut {NoStop}%
\bibitem [{\citenamefont {Nocera}\ \emph {et~al.}(2017)\citenamefont {Nocera},
  \citenamefont {Polkovnikov},\ and\ \citenamefont
  {Feiguin}}]{PhysRevA.95.023601}%
  \BibitemOpen
  \bibfield  {author} {\bibinfo {author} {\bibfnamefont {A.}~\bibnamefont
  {Nocera}}, \bibinfo {author} {\bibfnamefont {A.}~\bibnamefont
  {Polkovnikov}},\ and\ \bibinfo {author} {\bibfnamefont {A.~E.}\ \bibnamefont
  {Feiguin}},\ }\bibfield  {title} {\bibinfo {title} {Unconventional fermionic
  pairing states in a monochromatically tilted optical lattice},\ }\href
  {https://doi.org/10.1103/PhysRevA.95.023601} {\bibfield  {journal} {\bibinfo
  {journal} {Phys. Rev. A}\ }\textbf {\bibinfo {volume} {95}},\ \bibinfo
  {pages} {023601} (\bibinfo {year} {2017})}\BibitemShut {NoStop}%
\bibitem [{\citenamefont {Yu}\ \emph {et~al.}(2019)\citenamefont {Yu},
  \citenamefont {Jiang},\ and\ \citenamefont {Lu}}]{HHGreview}%
  \BibitemOpen
  \bibfield  {author} {\bibinfo {author} {\bibfnamefont {C.}~\bibnamefont
  {Yu}}, \bibinfo {author} {\bibfnamefont {S.}~\bibnamefont {Jiang}},\ and\
  \bibinfo {author} {\bibfnamefont {R.}~\bibnamefont {Lu}},\ }\bibfield
  {title} {\bibinfo {title} {High order harmonic generation in solids: a review
  on recent numerical methods},\ }\href
  {https://doi.org/10.1080/23746149.2018.1562982} {\bibfield  {journal}
  {\bibinfo  {journal} {Advances in Physics: X}\ }\textbf {\bibinfo {volume}
  {4}},\ \bibinfo {pages} {1562982} (\bibinfo {year} {2019})},\ \Eprint
  {https://arxiv.org/abs/https://doi.org/10.1080/23746149.2018.1562982}
  {https://doi.org/10.1080/23746149.2018.1562982} \BibitemShut {NoStop}%
\bibitem [{\citenamefont {Ghimire}\ and\ \citenamefont
  {Reis}(2018)}]{Ghimire2018}%
  \BibitemOpen
  \bibfield  {author} {\bibinfo {author} {\bibfnamefont {S.}~\bibnamefont
  {Ghimire}}\ and\ \bibinfo {author} {\bibfnamefont {D.~A.}\ \bibnamefont
  {Reis}},\ }\bibfield  {title} {\bibinfo {title} {High-harmonic generation
  from solids},\ }\href {https://doi.org/10.1038/s41567-018-0315-5} {\bibfield
  {journal} {\bibinfo  {journal} {Nat. Phys.}\ }\textbf {\bibinfo {volume}
  {15}},\ \bibinfo {pages} {10} (\bibinfo {year} {2018})}\BibitemShut {NoStop}%
\bibitem [{\citenamefont {McDonald}\ \emph {et~al.}(2015)\citenamefont
  {McDonald}, \citenamefont {Vampa}, \citenamefont {Orlando}, \citenamefont
  {Corkum},\ and\ \citenamefont {Brabec}}]{McDonald_2015}%
  \BibitemOpen
  \bibfield  {author} {\bibinfo {author} {\bibfnamefont {C.~R.}\ \bibnamefont
  {McDonald}}, \bibinfo {author} {\bibfnamefont {G.}~\bibnamefont {Vampa}},
  \bibinfo {author} {\bibfnamefont {G.}~\bibnamefont {Orlando}}, \bibinfo
  {author} {\bibfnamefont {P.~B.}\ \bibnamefont {Corkum}},\ and\ \bibinfo
  {author} {\bibfnamefont {T.}~\bibnamefont {Brabec}},\ }\bibfield  {title}
  {\bibinfo {title} {Theory of high-harmonic generation in solids},\ }\href
  {https://doi.org/10.1088/1742-6596/594/1/012021} {\bibfield  {journal}
  {\bibinfo  {journal} {Journal of Physics: Conference Series}\ }\textbf
  {\bibinfo {volume} {594}},\ \bibinfo {pages} {012021} (\bibinfo {year}
  {2015})}\BibitemShut {NoStop}%
\bibitem [{\citenamefont {Silva}\ \emph {et~al.}(2018)\citenamefont {Silva},
  \citenamefont {Blinov}, \citenamefont {Rubtsov}, \citenamefont {Smirnova},\
  and\ \citenamefont {Ivanov}}]{Silva2018}%
  \BibitemOpen
  \bibfield  {author} {\bibinfo {author} {\bibfnamefont {R.~E.}\ \bibnamefont
  {Silva}}, \bibinfo {author} {\bibfnamefont {I.~V.}\ \bibnamefont {Blinov}},
  \bibinfo {author} {\bibfnamefont {A.~N.}\ \bibnamefont {Rubtsov}}, \bibinfo
  {author} {\bibfnamefont {O.}~\bibnamefont {Smirnova}},\ and\ \bibinfo
  {author} {\bibfnamefont {M.}~\bibnamefont {Ivanov}},\ }\bibfield  {title}
  {\bibinfo {title} {{High-harmonic spectroscopy of ultrafast many-body
  dynamics in strongly correlated systems}},\ }\href@noop {} {\bibfield
  {journal} {\bibinfo  {journal} {Nature Photonics}\ }\textbf {\bibinfo
  {volume} {12}},\ \bibinfo {pages} {266} (\bibinfo {year} {2018})}\BibitemShut
  {NoStop}%
\bibitem [{sup()}]{supplement}%
  \BibitemOpen
  \href@noop {} {}\bibinfo {note} {See Supplemental Material at [URL will be
  inserted by publisher] for a full derivation of the ENZ field, a proof of the
  existence of the field, and the conditions under which it is guaranteed to be
  unique}\BibitemShut {NoStop}%
\bibitem [{\citenamefont {Pollack}\ \emph {et~al.}(2002)\citenamefont
  {Pollack}, \citenamefont {Pollack},\ and\ \citenamefont
  {Stump}}]{pollack2002electromagnetism}%
  \BibitemOpen
  \bibfield  {author} {\bibinfo {author} {\bibfnamefont {G.}~\bibnamefont
  {Pollack}}, \bibinfo {author} {\bibfnamefont {G.}~\bibnamefont {Pollack}},\
  and\ \bibinfo {author} {\bibfnamefont {D.}~\bibnamefont {Stump}},\
  }\href@noop {} {\emph {\bibinfo {title} {Electromagnetism}}}\ (\bibinfo
  {publisher} {Addison Wesley},\ \bibinfo {year} {2002})\BibitemShut {NoStop}%
\bibitem [{\citenamefont {Masur}\ \emph {et~al.}(2022)\citenamefont {Masur},
  \citenamefont {Bondar},\ and\ \citenamefont {McCaul}}]{PhysRevA.106.013110}%
  \BibitemOpen
  \bibfield  {author} {\bibinfo {author} {\bibfnamefont {J.}~\bibnamefont
  {Masur}}, \bibinfo {author} {\bibfnamefont {D.~I.}\ \bibnamefont {Bondar}},\
  and\ \bibinfo {author} {\bibfnamefont {G.}~\bibnamefont {McCaul}},\
  }\bibfield  {title} {\bibinfo {title} {Optical distinguishability of mott
  insulators in the time versus frequency domain},\ }\href
  {https://doi.org/10.1103/PhysRevA.106.013110} {\bibfield  {journal} {\bibinfo
   {journal} {Phys. Rev. A}\ }\textbf {\bibinfo {volume} {106}},\ \bibinfo
  {pages} {013110} (\bibinfo {year} {2022})}\BibitemShut {NoStop}%
\bibitem [{\citenamefont {Weinberg}\ and\ \citenamefont
  {Bukov}(2019)}]{quspin}%
  \BibitemOpen
  \bibfield  {author} {\bibinfo {author} {\bibfnamefont {P.}~\bibnamefont
  {Weinberg}}\ and\ \bibinfo {author} {\bibfnamefont {M.}~\bibnamefont
  {Bukov}},\ }\bibfield  {title} {\bibinfo {title} {{QuSpin: a Python Package
  for Dynamics and Exact Diagonalisation of Quantum Many Body Systems. Part II:
  bosons, fermions and higher spins}},\ }\href
  {https://doi.org/10.21468/SciPostPhys.7.2.020} {\bibfield  {journal}
  {\bibinfo  {journal} {SciPost Phys.}\ }\textbf {\bibinfo {volume} {7}},\
  \bibinfo {pages} {20} (\bibinfo {year} {2019})}\BibitemShut {NoStop}%
\bibitem [{\citenamefont {Hairer}\ \emph {et~al.}(1993)\citenamefont {Hairer},
  \citenamefont {Norsett},\ and\ \citenamefont {Wanner}}]{dop853}%
  \BibitemOpen
  \bibfield  {author} {\bibinfo {author} {\bibfnamefont {E.}~\bibnamefont
  {Hairer}}, \bibinfo {author} {\bibfnamefont {S.~P.}\ \bibnamefont
  {Norsett}},\ and\ \bibinfo {author} {\bibfnamefont {G.}~\bibnamefont
  {Wanner}},\ }\href {https://archive-ouverte.unige.ch/unige:12346} {\emph
  {\bibinfo {title} {Solving Ordinary Differential Equations I. Nonstiff
  Problems}}},\ \bibinfo {edition} {2nd}\ ed.\ (\bibinfo  {publisher}
  {Springer},\ \bibinfo {address} {Berlin},\ \bibinfo {year}
  {1993})\BibitemShut {NoStop}%
\bibitem [{\citenamefont {Virtanen}\ \emph {et~al.}(2020)\citenamefont
  {Virtanen}, \citenamefont {Gommers}, \citenamefont {Oliphant}, \citenamefont
  {Haberland}, \citenamefont {Reddy}, \citenamefont {Cournapeau}, \citenamefont
  {Burovski}, \citenamefont {Peterson}, \citenamefont {Weckesser},
  \citenamefont {Bright}, \citenamefont {{van der Walt}}, \citenamefont
  {Brett}, \citenamefont {Wilson}, \citenamefont {Millman}, \citenamefont
  {Mayorov}, \citenamefont {Nelson}, \citenamefont {Jones}, \citenamefont
  {Kern}, \citenamefont {Larson}, \citenamefont {Carey}, \citenamefont {Polat},
  \citenamefont {Feng}, \citenamefont {Moore}, \citenamefont {{VanderPlas}},
  \citenamefont {Laxalde}, \citenamefont {Perktold}, \citenamefont {Cimrman},
  \citenamefont {Henriksen}, \citenamefont {Quintero}, \citenamefont {Harris},
  \citenamefont {Archibald}, \citenamefont {Ribeiro}, \citenamefont
  {Pedregosa}, \citenamefont {{van Mulbregt}},\ and\ \citenamefont {{SciPy 1.0
  Contributors}}}]{2020SciPy-NMeth}%
  \BibitemOpen
  \bibfield  {author} {\bibinfo {author} {\bibfnamefont {P.}~\bibnamefont
  {Virtanen}}, \bibinfo {author} {\bibfnamefont {R.}~\bibnamefont {Gommers}},
  \bibinfo {author} {\bibfnamefont {T.~E.}\ \bibnamefont {Oliphant}}, \bibinfo
  {author} {\bibfnamefont {M.}~\bibnamefont {Haberland}}, \bibinfo {author}
  {\bibfnamefont {T.}~\bibnamefont {Reddy}}, \bibinfo {author} {\bibfnamefont
  {D.}~\bibnamefont {Cournapeau}}, \bibinfo {author} {\bibfnamefont
  {E.}~\bibnamefont {Burovski}}, \bibinfo {author} {\bibfnamefont
  {P.}~\bibnamefont {Peterson}}, \bibinfo {author} {\bibfnamefont
  {W.}~\bibnamefont {Weckesser}}, \bibinfo {author} {\bibfnamefont
  {J.}~\bibnamefont {Bright}}, \bibinfo {author} {\bibfnamefont {S.~J.}\
  \bibnamefont {{van der Walt}}}, \bibinfo {author} {\bibfnamefont
  {M.}~\bibnamefont {Brett}}, \bibinfo {author} {\bibfnamefont
  {J.}~\bibnamefont {Wilson}}, \bibinfo {author} {\bibfnamefont {K.~J.}\
  \bibnamefont {Millman}}, \bibinfo {author} {\bibfnamefont {N.}~\bibnamefont
  {Mayorov}}, \bibinfo {author} {\bibfnamefont {A.~R.~J.}\ \bibnamefont
  {Nelson}}, \bibinfo {author} {\bibfnamefont {E.}~\bibnamefont {Jones}},
  \bibinfo {author} {\bibfnamefont {R.}~\bibnamefont {Kern}}, \bibinfo {author}
  {\bibfnamefont {E.}~\bibnamefont {Larson}}, \bibinfo {author} {\bibfnamefont
  {C.~J.}\ \bibnamefont {Carey}}, \bibinfo {author} {\bibfnamefont
  {{\.I}.}~\bibnamefont {Polat}}, \bibinfo {author} {\bibfnamefont
  {Y.}~\bibnamefont {Feng}}, \bibinfo {author} {\bibfnamefont {E.~W.}\
  \bibnamefont {Moore}}, \bibinfo {author} {\bibfnamefont {J.}~\bibnamefont
  {{VanderPlas}}}, \bibinfo {author} {\bibfnamefont {D.}~\bibnamefont
  {Laxalde}}, \bibinfo {author} {\bibfnamefont {J.}~\bibnamefont {Perktold}},
  \bibinfo {author} {\bibfnamefont {R.}~\bibnamefont {Cimrman}}, \bibinfo
  {author} {\bibfnamefont {I.}~\bibnamefont {Henriksen}}, \bibinfo {author}
  {\bibfnamefont {E.~A.}\ \bibnamefont {Quintero}}, \bibinfo {author}
  {\bibfnamefont {C.~R.}\ \bibnamefont {Harris}}, \bibinfo {author}
  {\bibfnamefont {A.~M.}\ \bibnamefont {Archibald}}, \bibinfo {author}
  {\bibfnamefont {A.~H.}\ \bibnamefont {Ribeiro}}, \bibinfo {author}
  {\bibfnamefont {F.}~\bibnamefont {Pedregosa}}, \bibinfo {author}
  {\bibfnamefont {P.}~\bibnamefont {{van Mulbregt}}},\ and\ \bibinfo {author}
  {\bibnamefont {{SciPy 1.0 Contributors}}},\ }\bibfield  {title} {\bibinfo
  {title} {{{SciPy} 1.0: Fundamental Algorithms for Scientific Computing in
  Python}},\ }\href {https://doi.org/10.1038/s41592-019-0686-2} {\bibfield
  {journal} {\bibinfo  {journal} {Nature Methods}\ }\textbf {\bibinfo {volume}
  {17}},\ \bibinfo {pages} {261} (\bibinfo {year} {2020})}\BibitemShut
  {NoStop}%
\bibitem [{\citenamefont {McCaul}\ \emph {et~al.}(2023)\citenamefont {McCaul},
  \citenamefont {Jacobs},\ and\ \citenamefont {Bondar}}]{singleatom}%
  \BibitemOpen
  \bibfield  {author} {\bibinfo {author} {\bibfnamefont {G.}~\bibnamefont
  {McCaul}}, \bibinfo {author} {\bibfnamefont {K.}~\bibnamefont {Jacobs}},\
  and\ \bibinfo {author} {\bibfnamefont {D.~I.}\ \bibnamefont {Bondar}},\
  }\bibfield  {title} {\bibinfo {title} {Towards single atom computing via high
  harmonic generation},\ }\href@noop {} {\bibfield  {journal} {\bibinfo
  {journal} {EPJ PLus (accepted)}\ } (\bibinfo {year} {2023})}\BibitemShut
  {NoStop}%
\bibitem [{\citenamefont {Eeckhout}(2017)}]{eeckhout2017moore}%
  \BibitemOpen
  \bibfield  {author} {\bibinfo {author} {\bibfnamefont {L.}~\bibnamefont
  {Eeckhout}},\ }\bibfield  {title} {\bibinfo {title} {Is moore’s law slowing
  down? what’s next?},\ }\href@noop {} {\bibfield  {journal} {\bibinfo
  {journal} {IEEE Micro}\ }\textbf {\bibinfo {volume} {37}},\ \bibinfo {pages}
  {4} (\bibinfo {year} {2017})}\BibitemShut {NoStop}%
\end{thebibliography}%

\pagebreak
\widetext
\begin{center}
\textbf{\large Supplemental Materials for ``Dynamical Generation of Epsilon-Near-Zero Behaviour via Tracking and Feedback Control''}
\end{center}
\setcounter{equation}{0}
\setcounter{figure}{0}
\setcounter{table}{0}
\setcounter{page}{1}
\makeatletter
\renewcommand{\theequation}{S\arabic{equation}}
\renewcommand{\thefigure}{S\arabic{figure}}
\renewcommand{\bibnumfmt}[1]{[S#1]}
\renewcommand{\citenumfont}[1]{S#1}

\section{I. \ \ \ Derivation of the ENZ Field Determined by Tracking Control}
We begin our derivation from the condition for an ENZ-like response:
\begin{equation} \label{sup_enzcriteria}
    J_{\rm ENZ}(t) = -\frac{1}{a\ind} \Phi_{\rm ENZ}(t) + C.
\end{equation}
Note that as $\Phi(t)$ is proportional to the time integral of the field  $E_{\mathrm{in}} (t)$, we require $ \Phi(0)=0$, which fixes $C=J(0)$.

In order to perform tracking control, it is necessary to invert the expectation of the current operator
\begin{equation} \label{sup_currentop}
    \hat{J}(t) = -iat_0 \sum_{j, \sigma} \left(e^{-i\Phi(t)} \hat{c}^\dag_{j, \sigma} \hat{c}_{j+1, \sigma} - \mathrm{h.c.}\right)
\end{equation}
in order to express the control field in terms of $J(t)$, the expectation to be tracked. If the tracking condition is fulfilled at a time $t$ (which will be identically true at $t=0$), at time $t + \dt$, we require:
\begin{equation} \label{inductivestep}
    \langle \psi(t + \dt) | \hat{J}(t + \dt) | \psi(t + \dt) \rangle = -\frac{\Phi_{\rm ENZ}(t + \dt)}{a\ind} + J(0).
\end{equation}

In order to evaluate Eq.~\eqref{inductivestep}, we obtain the state of the system at time $t + \dt$ by the first order approximation to the Schr\"{o}dinger equation:
\begin{equation} \label{psiplusdt}
    |\psi(t + \dt)\rangle = |\psi(t)\rangle - i \dt \hat{H}(t) |\psi(t)\rangle + O(\dt^2).
\end{equation}
Substituting this into Eq.~\eqref{inductivestep} gives
\begin{equation} \label{firstorderexpansion}
        -\frac{\Phi_{\rm ENZ}(t + \dt)}{a\ind} + J(0) = \langle \psi(t) | \hat{J}(t+\dt) | \psi(t)\rangle
        + i\dt \langle \psi(t)| [\hat{H}(t), \hat{J}(t+\dt)] |\psi(t)\rangle + O(\dt^2).
\end{equation}

The commutator in the second term of the RHS can be expanded to
\begin{equation} \label{hamcurrentcomm}
        [\hat{H}(t), \hat{J}(t+\dt)] = iat_0\left\{ e^{-i\Phi_{\rm ENZ}(t+dt)} \left( t_0 e^{i\Phi_{\rm ENZ}(t))} [ \hat{K}^\dag, \hat{K} ]  \right.\right.
		\left.\left. - [ \hat{U}, \hat{K}]\right) +  e^{i\Phi_{\rm ENZ}(t+dt)} \left([ \hat{U}, \hat{K}^\dag ]-t_0e^{-i\Phi_{\rm ENZ}(t)} [ \hat{K}, \hat{K}^\dag ]\right)\right\},
\end{equation}
where for convenience we have defined the nearest neighbor operator $\hat{K}$ as
\begin{equation}
    \hat{K} = \sum_{j,\sigma} \hat{c}^\dag_{j, \sigma} \hat{c}_{j+1, \sigma}.
\end{equation}
Under periodic boundary conditions (which we assume throughout) $[\hat{K}, \hat{K}^\dag]=0$, meaning that it is possible rewrite Eq.~\eqref{inductivestep} by inserting Eq.~\eqref{hamcurrentcomm}, together with the definition of the current operator given by Eq.~\eqref{sup_currentop}:
\begin{equation} \label{interstep}
        -\frac{\Phi_{\rm ENZ}(t + \dt)}{a\ind} + J(0) = -iat_0 e^{-i\Phi_{\rm ENZ}(t+dt)} \left\{ \langle \psi | \hat{K} | \psi \rangle \right. \\
        \left. + i\dt \langle \psi | [\hat{U}, \hat{K}] | \psi \rangle \right\} + \mathrm{h.c.} + O(\dt^2)
\end{equation}
with the definition $|\psi\rangle \equiv |\psi(t)\rangle$.

This expression is further simplified by representing the expectation values in polar form as follows:
\begin{align}
	 \langle\psi|\hat{K}|\psi\rangle = \langle\psi|\hat{K}^\dag |\psi\rangle^\dag &= R(\psi) e^{i\theta(\psi)} \label{nnexpec} \\
	 \langle \psi | [\hat{U}, \hat{K}] | \psi \rangle = -\langle \psi | [\hat{U}, \hat{K}^\dag] | \psi \rangle^\dag &= P(\psi) e^{i\lambda(\psi)} \label{nnintcommexpec}
\end{align}
where we use the argument $\psi$ to indicate that the parameter is a functional of the state of the system $|\psi(t)\rangle$.

Thus, \eqref{interstep} becomes
\begin{equation}
    \begin{split}
        -\frac{\Phi_{\rm ENZ}(t+\dt)}{a\ind} + J(0) = -2at_0 R(\psi) \sin\left[\Phi_{\rm ENZ}(t+\dt) - \theta(\psi)\right] \\
        + 2at_0 P(\psi) \cos\left[\Phi_{\rm ENZ}(t+\dt) - \lambda(\psi)\right]\dt  + O(\dt^2).
	\end{split}
\end{equation}
Since the system in question is finite, each term in the above equation is bounded, so we can take the limit $\dt\rightarrow 0$ to obtain an implicit equation for $\Phi_{\rm ENZ}$:
\begin{equation} \label{sup_enzphi}
    \begin{split}
        \frac{1}{a\ind} \Phi_{\rm ENZ}(\psi) = 2at_0 R(\psi) \sin\left[ \Phi_{\rm ENZ}(\psi) - \theta(\psi) \right]+ J(0)
    \end{split}
\end{equation}
where we have replaced the argument of $\Phi_{\rm ENZ}$ with $\psi$ as all other variables are functionals of $\psi$, and hence the time dependence of $\Phi_{\rm ENZ}$ enters solely through the state of the system $|\psi(t)\rangle$. The solution to this implicit equation corresponds to the the laser field $-\frac{1}{a} d\Phi_{\rm ENZ}(t)/dt$ that will induce an ENZ response.

\section{II. \ \ \ Existence and Uniqueness of the ENZ Field}
Before using \eqref{sup_enzphi} to induce an ENZ response, it is important to consider under what conditions such a field exists. It is useful to recast the problem such that, instead of finding a solution, $\Phi(\psi)$, to \eqref{sup_enzphi}, we find the field for which
\begin{equation} \label{enz_zeros}
	f_\psi (\Phi) = \sin\left[\Phi - \theta(\psi) \right] - \frac{\Phi}{Y(\psi)} + G(\psi)
\end{equation}
has a zero. Where, for simplicity, we have defined $G(\psi) = \frac{R(\psi_0)}{R(\psi)}\sin[\theta(\psi_0)]$ and $Y(\psi) = 2a^2t_0 R(\psi)\ind$. Note that we treat $\Phi$ as a scalar parameter to this function, and the constants are entirely determined by the current state of the system.

\textbf{Theorem 1} - \textit{If $\Phi$ is a solution to \eqref{enz_zeros}, then $\Phi - Y(\psi) G(\psi)$ lies within the interval $\left[-|Y(\psi)|, |Y(\psi)| \right]$.}

\noindent \textit{Proof:} $\left|\sin\left[\Phi - \theta\right] \right| \leq 1$, so any solution must obey
\begin{equation}
    \left|\frac{\Phi}{Y(\psi)} - G(\psi) \right| \leq 1.
\end{equation}
It follows that
\begin{equation}
    -|Y(\psi)| \leq \Phi - Y(\psi)G(\psi) \leq |Y(\psi)| \square.
\end{equation}

\textbf{Theorem 2} - \textit{At least one solution to \eqref{enz_zeros} exists.}

\noindent \textit{Proof:} First, we evaluate the value of \eqref{enz_zeros} at the endpoints of the interval in which all possible solutions lie. Using $\xi_1 = -Y(\psi) + Y(\psi)G(\psi)$ and $\xi_2 = Y(\psi) + Y(\psi)G(\psi)$ the value of the function \eqref{enz_zeros} at the two endpoints of the solution interval is
\begin{align}
	f_\psi (\xi_1) &= \sin\left[ \xi_1 - \theta(\psi) \right] + 1 \geq 0\label{xi1}\\
	f_\psi (\xi_2) &= \sin\left[ \xi_2 - \theta(\psi) \right] - 1 \leq 0\label{xi2}.
\end{align} 
If at least one of $f_\psi(\xi_1)$ and $f_\psi(\xi_2)$ is zero, then there is at least one solution at one (or both) of the endpoints. If both are non-zero, then $f_\psi(\xi_1)$ is positive and $f_\psi(\xi_2)$ is negative, and since $f$ is continuous, there must be at least one zero in the interval $\left[-|Y(\psi)| + Y(\psi)G(\psi), |Y(\psi)| + Y(\psi)G(\psi) \right] \square$.

\textbf{Theorem 3} - \textit{If $|Y(\psi)| \leq 1$, there exists a unique solution to \eqref{sup_enzphi}.}

\noindent \textit{Proof:} The derivative of \eqref{enz_zeros} is 
\begin{equation}
	f'_\psi(\Phi) = \cos\left[ \Phi - \theta(\psi) \right] - \frac{1}{Y(\psi)}
\end{equation}
Hence, if $|Y(\psi)| \leq 1$, $f'(\Phi)$ is non-positive/negative and therefore $f(\Phi)$ is monotonic non-increasing/decreasing. Since $f_\psi(\Phi) = 0$ for some $\Phi$ by Theorem 2, this must be the only value for which $f_\psi(\Phi)$ obtains zero, and hence it is the unique solution to \eqref{sup_enzphi} $\square$. Note that since $R(\psi) \leq N_s$, the ENZ field is guaranteed to be unique regardless of the current state when $ \left|2a^2t_0 N_s \ind \right| \leq 1$.

\end{document}